%% file: main.tex
\documentclass[%
 sor,
 jor,
 amsmath,amssymb,
 reprint,%
author-numerical,%
]{revtex4-2}

\usepackage{graphicx}
\usepackage[caption=false]{subfig}
\renewcommand{\thesubfigure}{(\alph{subfigure})}
\newsavebox{\panelbox}
\newcommand{\panel}[2]{
  \subfloat[\label{#1}]{%
    \sbox{\panelbox}{#2}%
    \usebox{\panelbox}\hspace{-\wd\panelbox}%
    \rlap{\raisebox{\ht\panelbox}{\small\thesubfigure}}%
    \hspace{\wd\panelbox}}}
\usepackage{dcolumn}
\usepackage{bm}
\usepackage[acronym]{glossaries}
\makeatletter\def\active@comma{,}\makeatother

\begin{document}

\title{Analytic Two-Radius Model for the Zero-Shear and High-Frequency Viscosity of Colloidal Suspensions}

\author{Dimos Aslanis}
\affiliation{Institute for Theoretical Physics, Utrecht University,\\ Princetonplein 5, 3584 CC Utrecht, The Netherlands}
\affiliation{InProcess-LSP, Kloosterstraat 9, 5349 AB Oss, The Netherlands}

\author{Joost de Graaf}
\email[Author to whom correspondence should be addressed; electronic mail: ]{j.degraaf@uu.nl}
\affiliation{Institute for Theoretical Physics, Utrecht University,\\ Princetonplein 5, 3584 CC Utrecht, The Netherlands}

\begin{abstract}
Nearly hard-sphere colloids are stabilized by short-range repulsions that set their effective size and equilibrium microstructure.
Their hydrodynamic interactions, in contrast, are set by the no-slip surface. 
Since these two do not necessarily coincide, there are potentially two effective radii that control their dynamics in different ways.
Yet it is commonplace to model and analyze such suspensions with a single effective radius, which merges these different physics.
Two-radius models, in which a hard, solvent-permeable shell surrounds a no-slip core, have been proposed to improve the description of these suspensions. Their rheology is known exactly at pair level, but at higher concentrations their zero-shear viscosity has been obtained only with semi-empirical expressions, which do not recover the exact pair result.
Here, we close this gap with a theory for the zero-shear and high-frequency viscosity of these two-radius models.
By keeping pairs out of near contact, the shell lowers the high-frequency viscosity, as our Stokesian dynamics simulations confirm, and shifts the zero-shear viscosity from hydrodynamic toward thermodynamic stress.
We apply our theory to experiments on charge- and sterically stabilized colloids from the literature and find that it accurately describes the measured viscosities.
We discuss the benefits of fitting data to a two-radius model compared to a single-radius one.
To facilitate this fitting in future work, closed-form expressions are provided that cover the full range of shell thicknesses and volume fractions.
\end{abstract}

\maketitle

\input{sections/glossary}

\input{sections/introduction}
\input{sections/methods}
\input{sections/results_and_discussion}
\input{sections/conclusions}
\input{sections/acknowledgments}
\input{sections/declarations}

\input{sections/appendix}

\bibliographystyle{aipnum4-2}
\bibliography{references}

\end{document}

%% file: sections/glossary.tex
\newacronym[shortplural={SDs}, longplural={Stokesian dynamics}]{sd}{SD}{Stokesian dynamics}
\newacronym[shortplural={FSDs}, longplural={fast Stokesian dynamics}]{fsd}{FSD}{fast Stokesian dynamics}
\newacronym[shortplural={BDs}, longplural={Brownian dynamics}]{bd}{BD}{Brownian dynamics}
\newacronym[shortplural={GPUs}, longplural={Graphics Processing Units}]{gpu}{GPU}{Graphics Processing Unit}
\newacronym[shortplural={HSs}, longplural={Hard Spheres}]{hs}{HS}{Hard Sphere}
\newacronym[shortplural={HIs}, longplural={Hydrodynamic Interactions}]{hi}{HI}{Hydrodynamic Interaction}
\newacronym[shortplural={BHs}, longplural={Barker-Henderson models}]{bh}{BH}{Barker-Henderson}
\newacronym[shortplural={EVs}, longplural={Excluded Volumes}]{ev}{EV}{Excluded Volume}
\newacronym[shortplural={RPYs}, longplural={Rotne-Prager-Yamakawa tensors}]{rpy}{RPY}{Rotne-Prager-Yamakawa}
\newacronym[shortplural={PSEs}, longplural={Positively Split Ewald methods}]{pse}{PSE}{Positively Split Ewald}
\newacronym[shortplural={AdGs}, longplural={Alexander--de Gennes}]{adg}{AdG}{Alexander--de Gennes}
\newacronym[shortplural={MCTs}, longplural={mode-coupling theories}]{mct}{MCT}{mode-coupling theory}
\newacronym[shortplural={FIREs}, longplural={Fast Inertial Relaxation Engine}]{fire}{FIRE}{Fast Inertial Relaxation Engine}

%% file: sections/introduction.tex
\section{Introduction}
\label{sec:introduction}

Colloidal suspensions are found both in nature and in industrial products
and processes, ranging from blood and milk to paints, inks, coatings and
ceramics~\cite{russelColloidalDispersions1989,mewisColloidalSuspensionRheology2011}.
Hard spheres, which cannot overlap and otherwise do not interact, are a simple model for colloidal suspensions with steep, short-ranged
repulsions~\cite{mewisColloidalSuspensionRheology2011,royallColloidalHardSpheres2024}.
Such a repulsion can be mapped onto hard spheres with an effective diameter~\cite{barkerPerturbationTheoryEquation1967,royallSearchColloidalHard2013}.
The hard-sphere fluid has been studied extensively, both with
analytical
theories~\cite{thieleEquationStateHard1963,wertheimExactSolutionPercusYevick1963,carnahanEquationStateNonattracting1969,hansenTheorySimpleLiquids2013}
and with computer
simulations~\cite{alderPhaseTransitionHard1957,woodPreliminaryResultsRecalculation1957}.
Despite the absence of any attraction, hard spheres have a rich phase
behavior. As the volume fraction increases, the
fluid freezes into a crystal through a region of fluid--crystal
coexistence~\cite{alderPhaseTransitionHard1957,woodPreliminaryResultsRecalculation1957,hooverMeltingTransitionCommunal1968,fernandezEquilibriumFluidSolidCoexistence2012,puseyPhaseBehaviourConcentrated1986}.
If crystallization is avoided, for example by compression or by size
polydispersity, the dense fluid instead arrests into a
glass~\cite{woodcockGlassTransitionHardsphere1976,puseyObservationGlassTransition1987,puseyHardSpheresCrystallization2009}.
Real colloidal suspensions can be prepared that closely match hard spheres
in their equilibrium structure and phase
behavior~\cite{puseyPhaseBehaviourConcentrated1986,royallSearchColloidalHard2013}.
Because their particles are much larger and slower than atoms, their
structure and dynamics can be measured with static and dynamic light
scattering~\cite{puseyPhaseBehaviourConcentrated1986,puseyObservationGlassTransition1987}
and followed particle by particle with video and confocal
microscopy~\cite{crockerMethodsDigitalVideo1996,vanblaaderenRealSpaceStructureColloidal1995,weeksThreeDimensionalDirectImaging2000,prasadConfocalMicroscopyColloids2007}.
This makes colloidal hard spheres convenient experimental model systems for studying
liquids, crystals and
glasses~\cite{poonColloidsBigAtoms2004,royallColloidalHardSpheres2024}.

Hard spheres play a similar role in suspension rheology%
~\cite{bradyRheologicalBehaviorConcentrated1993,
lionbergerMicroscopicTheoriesRheology1999,mewisColloidalSuspensionRheology2011}.
Arguably, the most basic and widely measured rheological property is the
zero-shear viscosity $\eta_0$, that is, the viscosity in the limit of vanishing shear
rate, usually expressed relative to the solvent viscosity $\mu$.
The opposite limit of the linear viscoelastic spectrum is the high-frequency viscosity $\eta'_\infty$, which effectively is the viscosity of the suspension when it is oscillated too quickly for its microstructure to rearrange.
For hard
spheres, $\eta_0/\mu$ depends only on the particle volume fraction $\phi$, and
is known exactly to second order~\cite{einsteinNeueBestimmungMolekuldimensionen1906a,
einsteinBerichtigungMeinerArbeit1911,batchelorDeterminationBulkStress1972,
batchelorEffectBrownianMotion1977,cichockiLongtimeSelfdiffusionCoefficient1988}
in $\phi$. At higher volume fractions, $\eta_0/\mu$ rises steeply%
~\cite{dekruifHardSphereColloidal1985, vanderwerffHardsphereColloidalDispersions1989,
russelDivergenceLowShear2013a}, and theories rely on closures to describe the behavior~\cite{bradyRheologicalBehaviorConcentrated1993,
lionbergerMicroscopicTheoriesRheology1999,russelDivergenceLowShear2013a}.
Depending on the closure, $\eta_0/\mu$ diverges either at random close
packing~\cite{bradyRheologicalBehaviorConcentrated1993} or at the glass
transition~\cite{russelDivergenceLowShear2013a}, and which of the two applies
to real suspensions is still debated~\cite{chengNatureDivergenceLow2002,russelDivergenceLowShear2013a}.
Unlike the equilibrium properties of hard spheres, which depend only on
excluded volume, a purely geometric constraint, their rheology also depends
on the solvent-mediated hydrodynamic interactions between the
particles~\cite{fossBrownianDynamicsSimulation2000,fossStructureDiffusionRheology2000,royallColloidalHardSpheres2024}.
Hydrodynamic interactions require a boundary condition at the particle surface, and most theoretical treatments take the
particles to be rigid and impermeable, with smooth no-slip surfaces. Thus, a
single radius sets both the distance of closest approach and the surface at
which the solvent comes to rest relative to the
particle~\cite{russelDivergenceLowShear2013a}.

Realizing such particles experimentally is difficult, because colloids in general attract one another through van der Waals forces~\cite{russelColloidalDispersions1989}.
To behave as hard spheres, they must be kept apart by a steep, short-ranged repulsion~\cite{royallSearchColloidalHard2013}.
In practice, nearly hard-sphere behavior can be achieved with sterically stabilized
PMMA~\cite{antlPreparationPolymethylMethacrylate1986,puseyPhaseBehaviourConcentrated1986}
and silica~\cite{vanheldenPreparationCharacterizationSpherical1981,dekruifHardSphereColloidal1985}
particles, as well as with charge-stabilized particles in electrolyte solutions that
screen the electrostatic
repulsion~\cite{kriegerRheologyMonodisperseLatices1972}.
In each case, stabilization adds a repulsive layer at the particle surface.
This layer sets how closely particles approach and hence the equilibrium
microstructure, but it does not in general move with the particle as a rigid
shell. Solvent can partly flow through a polymer
layer~\cite{potaninHydrodynamicInteractionParticles1995}, and an
electrical double layer does not necessarily rotate with the
particle~\cite{kriegerRheologyMonodisperseLatices1972}. Abstracting small-scale detail, real particles can be seen to have an effective thermodynamic size and another, typically smaller one, when it comes to their hydrodynamic response. The
zero-shear and high-frequency viscosities are not equally sensitive to
structure and hydrodynamics,~\textit{i.e.}, they weigh the two sizes differently. This leads to an effective suspension viscosity that departs from the hard-sphere expectation.

In describing experiments on nearly hard-sphere systems with low polydispersity, the two sizes are usually treated as one. That is, the difference is considered small and the response of the suspension is mapped to a single effective particle size by introducing an effective volume fraction.
The effective volume fraction has been fitted to hydrodynamic properties, such as the intrinsic viscosity of dilute suspensions~\cite{mewisRheologySuspensionsContaining1989}, or to structural properties, such as the freezing transition~\cite{puseyPhaseBehaviourConcentrated1986,poonMeasuringColloidalVolume2012}.
In concentrated suspensions, it has also been chosen so that the viscosity matches hard-sphere results, for example by placing the divergence of $\eta_0$ at the hard-sphere glass transition~\cite{hornHydrodynamicColloidalInteractions2000} or by matching the high-shear or high-frequency viscosity to hard-sphere master curves~\cite{russelDivergenceLowShear2013a}.
The effective size obtained in this way depends on the property, and the range of $\phi$, used to determine it~\cite{poonMeasuringColloidalVolume2012}.

A first step toward a more realistic, but still simple, description is to separate
the two sizes. The two-radius (or excluded-shell, or excluded-annulus) model does this by
assigning a hydrodynamic radius $a$ to the no-slip surface and a larger radius
$b$ to hard-sphere exclusion, with the shell between the two radii fully permeable
to the solvent. This makes explicit the above abstraction. The reduction is meaningful as it has been shown to closely reproduce the hydrodynamics of
particles with an internal structure, such as particles with a thin permeable polymer shell~\cite{cichockiHydrodynamicRadiusApproximation2014}.
Most rheological applications of the model are limited to second order
in $\phi$, where only pair interactions are needed.
Russel~\cite{russelHugginsCoefficientMeans1984} used a far-field approximation of
the pair hydrodynamics to calculate the $\phi^2$ coefficient of the zero-shear
viscosity as a function of $b/a$. Bergenholtz
\textit{et al.}~\cite{bergenholtzNonNewtonianRheologyDilute2002} solved the same
pair problem numerically with the full pair hydrodynamics. They obtained $\eta_0$,
$\eta'_\infty$ and the shear-rate dependence of the viscosity and normal stress
differences. Swan \textit{et al.}~\cite{swanMediumAmplitudeOscillatory2014}
extended this to oscillatory shear and calculated the linear viscoelastic spectrum
between these two limits. The model has also been used for the microstructure of
strongly sheared suspensions~\cite{bradyMicrostructureStronglySheared1997},
microrheology~\cite{khairSingleParticleMotion2006} and the linear viscoelasticity of attractive
dispersions~\cite{vargaLinearViscoelasticityAttractive2015}.

Beyond the pair level, studies of two-radius particles have mainly addressed
short-time properties, including the high-frequency viscosity%
~\cite{abadeHighfrequencyViscosityConcentrated2010,abadeDiffusionSedimentationRheology2012,
riestDynamicsSuspensionsHydrodynamically2015} and the sedimentation and short-time
self-diffusion coefficients~\cite{abadeDiffusionSedimentationRheology2012,
riestDynamicsSuspensionsHydrodynamically2015}. The zero-shear viscosity of two-radius particles at finite concentrations has
also been studied, with approximate closures and semi-empirical expressions.
Brady~\cite{bradyRheologicalBehaviorConcentrated1993} derived a scaling
description for hard spheres and extended it to two-radius particles.
His model relates the thermodynamic contribution to the equilibrium pair
distribution at contact and the short-time self-diffusion coefficient.
However, after scaling out this diffusivity, he neglected hydrodynamic
interactions in calculating the flow-induced distortion of the microstructure.
Riest \textit{et al.}~\cite{riestDynamicsSuspensionsHydrodynamically2015} applied
Brady's hard-sphere expression to two-radius particles through semi-empirical
expressions for the short-time self-diffusivity and $\eta'_\infty$.
They replaced its contact-value factor by a fit to \gls{bd} data for
hard spheres without hydrodynamic interactions, which improved the prediction close to the divergence of $\eta_0$.
Mendoza~\cite{mendozaEffectiveStaticHighfrequency2011} took a different approach and
obtained $\eta_0$ and $\eta'_\infty$ for core--shell particles, which include
two-radius particles as a limit, through a differential effective-medium construction.
All three approaches have a theoretical basis and are capable of describing the experimental data quite well.
However, they depend partly on simulation data, either as tabulated input
or through fitted parameters. In addition, they do not recover the exact dependence
of the pair coefficient of the zero-shear viscosity on the shell thickness.

In this work, we develop a (semi-)analytic theory for both the zero-shear and the
high-frequency viscosity of two-radius hard spheres at finite concentrations.
We split the zero-shear viscosity into a hydrodynamic
contribution, equal to the high-frequency viscosity, and a thermodynamic contribution, due to
the distortion of the microstructure by the flow (Sec.~\ref{subsec:two_radius_model}). At pair level, the shell
enters these two parts through two functions of $b/a$, the integral $I$ of the
pair stresslet and the amplitude $\mathcal{A}$ of the thermodynamic viscosity.
We compute both exactly, using recent accurate methods for the pair hydrodynamic
interactions~\cite{wilsonStokesFlowThree2013,townsendGeneratingScratchNearfield2023} (Secs.~\ref{subsec:hydrodynamic_viscosity} and~\ref{subsec:thermodynamic_viscosity}).
We then extend the pair results to higher concentrations. For the hydrodynamic part, we use the self-consistent closure
of Potanin~\cite{potaninSelfConsistentCalculationsViscosity1993}, which we
test against \gls{sd} simulations (Sec.~\ref{subsec:hard_sphere_simulations}). For the thermodynamic part, we use
the \gls{mct}~\cite{russelDivergenceLowShear2013a} and
Brady~\cite{bradyRheologicalBehaviorConcentrated1993} closures (Sec.~\ref{subsec:thermodynamic_viscosity}). In both closures, we replace the approximate pair amplitude used so far
by the exact $\mathcal{A}$. Together with simple, accurate interpolations of $I$ and
$\mathcal{A}$, we get closed-form expressions for both viscosities at any shell
thickness and volume fraction, which can readily be used to analyze rheological data (Secs.~\ref{subsec:res_hydrodynamic}--\ref{subsec:res_closures}).
Finally, we compare the theory with measurements on charge-stabilized and
sterically stabilized colloids (Sec.~\ref{subsec:res_experiment}).

%% file: sections/methods.tex
\section{Methods}
\label{sec:methods}

We first introduce the two-radius model and split its zero-shear viscosity
into a hydrodynamic and a thermodynamic part. We then evaluate each part at
pair level and extend it to finite concentrations. Finally, we describe the
simulations used to test the hydrodynamic contribution to the viscosity.

\subsection{Two-radius hard-sphere model for the zero-shear viscosity}
\label{subsec:two_radius_model}

\begin{figure}[!tb]
  \includegraphics[scale=1]{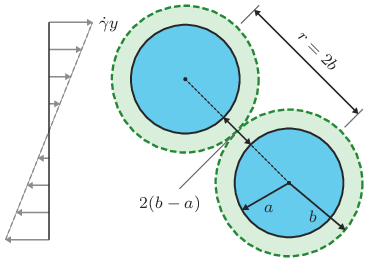}
  \caption{Two-radius model, with $\Lambda = b/a = 1.35$ in this example.
  Each particle has a no-slip core of hydrodynamic radius $a$ (cyan) and a hard shell of excluded-volume radius $b$ (green, dashed).
  The gray arrows show the imposed simple shear flow, $\mathbf{u}^{\infty}=\dot\gamma\,y\,\hat{\mathbf{x}}$, and the pair is at contact, $r = 2b$, along the compressional axis.
  }
  \label{fig:two_radius_model}
\end{figure}

We represent each particle by two concentric surfaces (Fig.~\ref{fig:two_radius_model}). The
\emph{hydrodynamic radius} $a$ sets the surface on which the no-slip boundary
condition is imposed in the Stokes-flow calculation.
The \emph{excluded-volume}, or \emph{thermodynamic}, radius
$b \ge a$ sets the distance of closest approach between particle centers.
The shell between the radii $a$ and $b$ represents a repulsive layer (steric, electrostatic,
or otherwise)~\cite{royallSearchColloidalHard2013,royallColloidalHardSpheres2024} that is permeable to the suspending medium.
To keep the model as simple and as general as possible we take the repulsion
to be hard, so that the pair potential at center-to-center separation $r$ is
\begin{equation}
  V(r)=
  \begin{cases}
    \infty, & r<2b,\\
    0,      & r\ge2b,
  \end{cases}
  \label{eq:V_shell}
\end{equation}
and the two radii enter only through the single ratio $\Lambda = b/a$.
This approach interpolates between two limits: at $\Lambda = 1$ a single radius sets both
surfaces and ideal hard spheres are recovered, while in the limit $\Lambda \to \infty$
pairs are held outside the range of their mutual hydrodynamic interaction and
the hard-sphere suspension without hydrodynamics is
recovered~\cite{bergenholtzNonNewtonianRheologyDilute2002}. The two radii give
rise to two volume fractions that determine the properties of the system.
(i) The thermodynamic volume fraction
$\phi_{b} = \tfrac{4}{3}\pi n b^{3}$, with $n$ the particle number density,
controls the equilibrium microstructure. (ii) The hydrodynamic volume fraction
$\phi_{a} = \Lambda^{-3}\phi_{b}$ is relevant for the fluid-mediated
interactions between particles.

On colloidal length and velocity scales the Reynolds number is negligible, so the solvent velocity $\mathbf{u}$ and
pressure $p$ satisfy the steady, incompressible Stokes
equations~\cite{kimMicrohydrodynamicsPrinciplesSelected2005,grahamMicrohydrodynamicsBrownianMotion2018}
\begin{equation}
  \nabla\cdot\boldsymbol{\sigma} = \mathbf{0},
  \qquad
  \nabla\cdot\mathbf{u} = 0,
  \qquad
  \boldsymbol{\sigma}
  = -p\,\mathbf{I} + \mu\big(\nabla\mathbf{u} + \nabla\mathbf{u}^{\mathsf{T}}\big).
  \label{eq:stokes}
\end{equation}
Here $\boldsymbol{\sigma}$ is the stress tensor, $\mu$ the dynamic viscosity
of the solvent and $\mathbf{I}$ the identity tensor. The first equation
expresses the balance of forces in the absence of inertia, the second
incompressibility, and the third the Newtonian constitutive relation.
Equation~\eqref{eq:stokes} is solved subject to no slip on the surface
of each particle, at a distance $a$ from its center, and to
$\mathbf{u} \to \mathbf{u}^{\infty}(\mathbf{r})=\langle\boldsymbol{\Gamma}\rangle\cdot\mathbf{r}$
far away. Here $\langle\boldsymbol{\Gamma}\rangle$ is the imposed velocity
gradient, whose symmetric part is the rate of strain $\langle\mathbf{E}\rangle$.
In simple shear with shear rate $\dot\gamma$,
$\mathbf{u}^{\infty}=\dot\gamma\,y\,\hat{\mathbf{x}}$, so that
$\langle\boldsymbol{\Gamma}\rangle=\dot\gamma\,\hat{\mathbf{x}}\hat{\mathbf{y}}$ and
$\langle\mathbf{E}\rangle=\tfrac12\dot\gamma
(\hat{\mathbf{x}}\hat{\mathbf{y}}+\hat{\mathbf{y}}\hat{\mathbf{x}})$. The excluded-volume
radius does not enter the boundary conditions of Eq.~\eqref{eq:stokes}
directly. It appears only through the configurations over which the particle
suspension is averaged.

Because Eq.~\eqref{eq:stokes} is linear, the hydrodynamic forces and torques
on the particles, collected in $\mathbf{F}$, and their stresslets $\mathbf{S}$
are linear in the particle translational and rotational velocities
$\mathbf{U}$ and the imposed
flow~\cite{kimMicrohydrodynamicsPrinciplesSelected2005,grahamMicrohydrodynamicsBrownianMotion2018},
\begin{equation}
  \begin{pmatrix} \mathbf{F} \\[2pt] \mathbf{S} \end{pmatrix}
  = -\underbrace{
  \begin{pmatrix}
     \mathbf{R}_{FU} & \mathbf{R}_{FE} \\[2pt]
     \mathbf{R}_{SU} & \mathbf{R}_{SE}
  \end{pmatrix}}_{\boldsymbol{\mathcal{R}}}
  \cdot
  \begin{pmatrix}
     \mathbf{U} - \mathbf{U}^{\infty} \\[2pt] -\langle \mathbf{E} \rangle
  \end{pmatrix},
  \label{eq:GRM}
\end{equation}
where $\boldsymbol{\mathcal{R}}$ is the grand resistance matrix and $\mathbf{U}^{\infty}$ contains the translational and rotational velocities of the imposed flow at the particle centers. If the
particles carry no external force or torque and their inertia is negligible, the hydrodynamic force is balanced
instantaneously by the interparticle force $\mathbf{F}^{P}$ and the Brownian
force $\mathbf{F}^{B}$, and the force row of Eq.~\eqref{eq:GRM} gives
\begin{equation}
  \mathbf{U} - \mathbf{U}^{\infty}
  = \mathbf{R}_{FU}^{-1}\cdot
  \big( \mathbf{R}_{FE}:\langle\mathbf{E}\rangle
        + \mathbf{F}^{P} + \mathbf{F}^{B} \big).
  \label{eq:Uforcebal}
\end{equation}
Inserting Eq.~\eqref{eq:Uforcebal} into the stresslet row of Eq.~\eqref{eq:GRM} splits the stresslet of each particle into a part proportional to $\langle\mathbf{E}\rangle$ and a part driven by the interparticle and Brownian forces.
Averaging this result over the Brownian fluctuations and over configurations, and adding the direct contribution of the interparticle forces and the isotropic kinetic stress $-nk_BT\mathbf{I}$, which does not contribute to the viscosity, gives the particle stress~\cite{bossisRheologyBrownianSuspensions1989,bradyBrownianMotionHydrodynamics1993}
\begin{equation}
  \big\langle\boldsymbol{\Sigma}_{p}\big\rangle
  = -\,nk_BT\,\mathbf{I}
  + n\Big[ \big\langle\mathbf{S}^{H}\big\rangle
         + \big\langle\mathbf{S}^{B}\big\rangle
         + \big\langle\mathbf{S}^{P}\big\rangle \Big],
  \label{eq:Sigma_p}
\end{equation}
with
\begin{align}
  \big\langle\mathbf{S}^{H}\big\rangle
  &= -\Big\langle \mathbf{R}_{SU}\cdot\mathbf{R}_{FU}^{-1}\cdot\mathbf{R}_{FE}
       - \mathbf{R}_{SE} \Big\rangle : \langle\mathbf{E}\rangle ,
  \label{eq:SH_def}\\[2pt]
  \big\langle\mathbf{S}^{B}\big\rangle
  &= -\,k_BT\,\Big\langle \nabla\cdot
       \big( \mathbf{R}_{SU}\cdot\mathbf{R}_{FU}^{-1} \big) \Big\rangle ,
  \label{eq:SB_def}\\[2pt]
  \big\langle\mathbf{S}^{P}\big\rangle
  &= -\,\Big\langle \big( \mathbf{x}\mathbf{I}
       + \mathbf{R}_{SU}\cdot\mathbf{R}_{FU}^{-1} \big)\cdot\mathbf{F}^{P}
     \Big\rangle ,
  \label{eq:SP_def}
\end{align}
where $\langle\cdot\rangle$ denotes an average over configuration space,
$\mathbf{x}$ is the particle position, $k_B$ is the Boltzmann constant and $T$
the temperature. Equations~\eqref{eq:SH_def}--\eqref{eq:SP_def} are the
stresslets induced by the imposed strain, by Brownian motion and by the
interparticle forces, respectively. We discuss how each enters the viscosity
below Eq.~\eqref{eq:eta_from_stress}. In the weak-flow limit the deviatoric bulk stress,
equal to the solvent stress $2\mu\langle\mathbf{E}\rangle$ plus the deviatoric part 
of Eq.~\eqref{eq:Sigma_p}, takes the Newtonian form $2\eta_0\langle\mathbf{E}\rangle$, which
defines the zero-shear viscosity $\eta_0$. In simple shear its shear component
is $\eta_0\dot\gamma$, so that
\begin{equation}
  \frac{\eta_{0}}{\mu}
  = 1 + \lim_{\dot\gamma\to0}\frac{n}{\mu\dot\gamma}
    \Big( \big\langle S^{H}_{xy}\big\rangle
        + \big\langle S^{B}_{xy}\big\rangle
        + \big\langle S^{P}_{xy}\big\rangle \Big).
  \label{eq:eta_from_stress}
\end{equation}

The three terms of Eq.~\eqref{eq:eta_from_stress} do not enter at the same
order in the flow-induced distortion of the microstructure.
$\langle\mathbf{S}^{H}\rangle$ is explicitly proportional to
$\langle\mathbf{E}\rangle$, so it survives averaging over the equilibrium
microstructure and yields a finite viscosity. Together with the solvent viscosity, this contribution gives
the high-frequency dynamic viscosity $\eta'_{\infty}$, in which the
structure has no time to respond to the
flow~\cite{bradyRheologicalBehaviorConcentrated1993}. By contrast,
$\langle\mathbf{S}^{B}\rangle$ and $\langle\mathbf{S}^{P}\rangle$ contain no
explicit rate of strain. Averaged over the isotropic equilibrium structure they
are purely isotropic and their deviatoric parts vanish identically~\cite{bradyBrownianMotionHydrodynamics1993,bergenholtzNonNewtonianRheologyDilute2002}.
Their deviatoric parts are nonzero only because the flow distorts the
microstructure, so each is of order $\dot\gamma$, and the division by
$\dot\gamma$ in Eq.~\eqref{eq:eta_from_stress} leaves a finite, rate-independent
viscosity.

Accordingly, we can write
\begin{equation}
  \frac{\eta_{0}}{\mu} = \frac{\eta'_{\infty}}{\mu} + \frac{\eta_{T}}{\mu},
  \qquad \eta_{T} = \eta_{B} + \eta_{P}.
  \label{eq:zero-shear_viscosity_total}
\end{equation}
Comparing with Eq.~\eqref{eq:eta_from_stress}, $\eta'_\infty/\mu = 1 + n\langle S^{H}_{xy}\rangle/(\mu\dot\gamma)$, while $\eta_B/\mu = n\langle S^{B}_{xy}\rangle/(\mu\dot\gamma)$ and $\eta_P/\mu = n\langle S^{P}_{xy}\rangle/(\mu\dot\gamma)$.
We call the sum of the last two, $\eta_T$, the thermodynamic viscosity.

\subsection{Hydrodynamic viscosity}
\label{subsec:hydrodynamic_viscosity}

For a suspension of identical rigid
spheres Batchelor and Green~\cite{batchelorDeterminationBulkStress1972}
evaluated the equilibrium average of the hydrodynamic stresslet,
Eq.~\eqref{eq:SH_def}, to $\mathcal{O}(\phi^{2})$, obtaining
\begin{equation}
  \frac{\eta'_\infty}{\mu}
  = 1 + \tfrac{5}{2}\phi_{a}
    + \phi_{a}^{2}\,\underbrace{\left[\, \tfrac{5}{2}
      + \tfrac{15}{2}\int_{2}^{\infty}\! g_{\mathrm{eq}}(as)\,J(s)\,s^{2}\,\mathrm{d}s \,\right]}_{c_H} + \mathrm{h.o.t.},
  \label{eq:BG_c2}
\end{equation}
with $s = r/a$ the center-to-center separation scaled on the hydrodynamic
radius, $g_{\mathrm{eq}}$ the pair distribution function and $J$ the stresslet function
discussed below. The volume fraction appearing here is the hydrodynamic one.
The linear coefficient in Eq.~\eqref{eq:BG_c2} is called the intrinsic viscosity,
$[\eta]=5/2$, and the product $[\eta]\phi_a$ is the Einstein term.
The quadratic coefficient $c_H$ is proportional to the Huggins coefficient%
~\cite{russelHugginsCoefficientMeans1984,doiTheoryPolymerDynamics2013},
$k_H = c_H/[\eta]^2$ (or $k_H = c_H/[\eta]$ in the definition of
Ref.~\onlinecite{swanMediumAmplitudeOscillatory2014}).
The Einstein term and the first term in the bracket, both with coefficient $5/2$,
are properties of the rigid, solvent-impermeable cores.
They are set by $\phi_{a}$ alone and do not depend on $\Lambda$.
In the dilute limit, the equilibrium pair distribution $g_{\mathrm{eq}}$ is the Boltzmann factor~\cite{russelColloidalDispersions1989}
associated with the potential in Eq.~\eqref{eq:V_shell},
\begin{equation}
  g_{\mathrm{eq}}(r) \approx e^{-V(r)/k_BT} = \Theta(r - 2b).
  \label{eq:pair_eq}
\end{equation}
Here $\Theta$ is the Heaviside step function, which excludes separations
$r<2b$. In the scaled variable of Eq.~\eqref{eq:BG_c2} this becomes
$g_{\mathrm{eq}}(as)=\Theta(s-2\Lambda)$, so the integral effectively
starts at $s=2\Lambda$ rather than at hard-core contact. Finite-concentration
corrections to this dilute pair distribution
are $\mathcal{O}(\phi_b)$ and first contribute to
the viscosity at cubic order in concentration~\cite{russelColloidalDispersions1989}.

To extend Eq.~\eqref{eq:BG_c2} beyond pair level we adopt the self-consistent
method of Potanin~\cite{potaninSelfConsistentCalculationsViscosity1993}, in
which each pair is taken to interact through the suspension rather than through
the pure solvent. The two-body hydrodynamic functions are left unchanged and
$\mu$ is replaced by the suspension viscosity only in the integral term of Eq.~\eqref{eq:BG_c2}, which carries the hydrodynamic interaction within a pair.
This multiplies that term by $\eta'_{\infty}/\mu$ and makes the result implicit in $\eta'_{\infty}$. Solving
for $\eta'_{\infty}$ gives
\begin{equation}
  \frac{\eta'_\infty}{\mu}
  = \frac{1 + \tfrac{5}{2}\,\phi_a + \tfrac{5}{2}\,\phi_a^{2}}
         {1 - \tfrac{15}{2}\,\phi_a^{2}\, I(\Lambda)},
  \qquad
  I(\Lambda) = \int_{2\Lambda}^{\infty} J(s)\, s^{2}\, \mathrm{d}s .
  \label{eq:eta_inf}
\end{equation}
Expanded to $\mathcal{O}(\phi_{a}^{2})$, Eq.~\eqref{eq:eta_inf} returns
$c_H=\tfrac{5}{2} + \tfrac{15}{2}I(\Lambda)$ for the quadratic coefficient, so the
closure reproduces the exact pair result for every $\Lambda$ by construction.
The excluded-volume radius enters only through the lower limit of
$I(\Lambda)$, as in earlier pair-level
calculations~\cite{russelHugginsCoefficientMeans1984,cichockiShorttimeDynamicsHighfrequency2013}.

The kernel $J(s)$ is the Batchelor--Green stresslet
function defined in Ref.~\onlinecite{batchelorDeterminationBulkStress1972}. In terms of the two-sphere
mobility functions~\cite{kimResistanceMobilityFunctions1985,
kimMicrohydrodynamicsPrinciplesSelected2005}, it is%
\footnote{In the convention used here, the stresslet is exerted by the fluid on the
particle. Kim and Mifflin~\cite{kimResistanceMobilityFunctions1985} take the
stresslet exerted by the particle on the fluid, so that their mobility
functions follow from ours by $x^m_1\to-x^m_1$, $y^m_1\to-y^m_1$, and
$z^m_1\to-z^m_1$, and Eq.~\eqref{eq:J_mobility} becomes
$J=-1-\tfrac15(x^m_1+2y^m_1+2z^m_1)$.}
\begin{equation}
  J(s) = -1 + \tfrac{1}{5}\left[
    x_1^m(s) + 2y_1^m(s) + 2z_1^m(s) \right],
  \label{eq:J_mobility}
\end{equation}
and it decays as $J \sim \tfrac{15}{2}s^{-6}$ for $s \gg 1$, with $J(2) = 0.2214$ at
contact~\cite{batchelorDeterminationBulkStress1972}. The mobility functions
$x_1^m$, $y_1^m$, and $z_1^m$ are lengthy and are not reproduced here. Details on their form can be found in
Refs.~\onlinecite{kimResistanceMobilityFunctions1985,kimMicrohydrodynamicsPrinciplesSelected2005}. We compute the mobility
functions from lubrication asymptotics near contact and a multipole-reflection
solution at larger separations, as described in Appendix~\ref{app:numerical-method}.

Quadrature of Eq.~\eqref{eq:eta_inf} gives $I(1) = 0.3336$, whence
$c_H(1)=5.00$, in agreement
with literature
values~\cite{cichockiShorttimeDiffusionCoefficients1988,bergenholtzNonNewtonianRheologyDilute2002}.
The denominator of Eq.~\eqref{eq:eta_inf} vanishes at
$\phi_a = \phi_\mathrm{div}(\Lambda) = [\tfrac{15}{2}I(\Lambda)]^{-1/2}$, and at
$\Lambda = 1$ this gives $\phi_\mathrm{div} = 0.632$, close to random close
packing of monodisperse hard spheres~\cite{scottDensityRandomClose1969,berrymanRandomClosePacking1983,torquatoJammedHardparticlePackings2010}.

\subsection{Thermodynamic viscosity}
\label{subsec:thermodynamic_viscosity}

Because the Brownian and interparticle stresses, Eqs.~\eqref{eq:SB_def} and
\eqref{eq:SP_def}, contribute only through the flow-induced distortion of the
pair distribution, evaluating $\eta_T$ requires the solution of the pair
Smoluchowski problem rather than an equilibrium average.

At pair level and under weak flows, the thermodynamic viscosity can be written
as~\cite{batchelorEffectBrownianMotion1977,bergenholtzNonNewtonianRheologyDilute2002}
\begin{equation}
  \frac{\eta_T}{\mu} = \mathcal{A}(\Lambda)\,\phi_b^{2} + \mathrm{h.o.t.},
  \label{eq:Adef}
\end{equation}
where the amplitude $\mathcal{A}$ collects the $\mathcal{O}(\phi_b^{2})$ contribution
from the Brownian and interparticle stresses.

Two limits of $\mathcal{A}$ are known. At $\Lambda = 1$ it reduces to Batchelor's
coefficient for ideal hard
spheres~\cite{batchelorEffectBrownianMotion1977}, subsequently
refined~\cite{cichockiLongtimeSelfdiffusionCoefficient1988} to
$\mathcal{A}(1) = 0.913$.
As $\Lambda \to \infty$ the hydrodynamic interactions within a pair become
negligible, the pair Smoluchowski problem can be solved analytically%
~\cite{russelHugginsCoefficientMeans1984,bradyRheologicalBehaviorConcentrated1993},
and $\mathcal{A} \to \tfrac{12}{5}\Lambda^{-1}$.
At intermediate $\Lambda$ values, $\mathcal{A}$ must be
computed numerically~\cite{bergenholtzNonNewtonianRheologyDilute2002}.
Therefore, we compute it for each $\Lambda$ from the full
two-particle hydrodynamics by solving the two-particle Smoluchowski equation
as described below.

We scale the separation on the excluded-volume radius, $\rho = r/b = s/\Lambda$,
so that the excluded shell lies at $\rho = 2$ for every $\Lambda$, and define
the excluded-volume P\'eclet number~\cite{bradyMicrostructureStronglySheared1997}
\begin{equation}
  \mathrm{Pe}_b = \frac{b^{2}\dot\gamma}{2D_0},
  \qquad D_0 = \frac{k_BT}{6\pi\mu a},
  \label{eq:Peb}
\end{equation}
where $D_0$ is the Stokes--Einstein diffusivity of an isolated particle.
When only pair interactions are retained, the steady pair distribution satisfies the
two-particle Smoluchowski
equation~\cite{batchelorEffectBrownianMotion1977,russelColloidalDispersions1989,bergenholtzNonNewtonianRheologyDilute2002}
\begin{equation}
  \nabla\cdot\left(\mathbf{D}_r\cdot\nabla g\right)
  = \mathrm{Pe}_b\,\nabla\cdot\left(\mathbf{U}_r\, g\right).
  \label{eq:smol}
\end{equation}
Here $\nabla$ is the gradient with respect to the dimensionless separation
$\boldsymbol{\rho}=\mathbf{r}/b$. The boundary conditions are vanishing radial
flux at $\rho=2$ and $g\to1$ as $\rho\to\infty$. The relative diffusivity and
velocity are~\cite{batchelorEffectBrownianMotion1977,bergenholtzNonNewtonianRheologyDilute2002}
\begin{align}
  \mathbf{D}_r &= G(s)\,\hat{\mathbf{r}}\hat{\mathbf{r}}
    + H(s)\left(\mathbf{I} - \hat{\mathbf{r}}\hat{\mathbf{r}}\right),
  \label{eq:Drel}\\
  \mathbf{U}_r &= \hat{\boldsymbol{\Gamma}}\cdot\boldsymbol{\rho}
    - \boldsymbol{\rho}\cdot\hat{\mathbf{E}}\cdot
      \left[\,A(s)\,\hat{\mathbf{r}}\hat{\mathbf{r}}
      + B(s)\left(\mathbf{I} - \hat{\mathbf{r}}\hat{\mathbf{r}}\right)\right]
  \nonumber\\
  &= \hat{\boldsymbol{\Omega}}\cdot\boldsymbol{\rho}
    + \boldsymbol{\rho}\cdot\hat{\mathbf{E}}\cdot
      \left\{\left[1-A(s)\right]\hat{\mathbf{r}}\hat{\mathbf{r}}
      + \left[1-B(s)\right]\left(\mathbf{I} - \hat{\mathbf{r}}\hat{\mathbf{r}}\right)\right\},
  \label{eq:Urel}
\end{align}
where $\mathbf{D}_r$ is scaled on the far-field relative diffusivity $2D_0$
and $\mathbf{U}_r$ on $b\dot\gamma$.
Here $\hat{\mathbf{r}}=\boldsymbol{\rho}/\rho$ is the unit vector along the line of centers.
The functions $G$ and $H$ are the relative mobilities parallel and transverse to the
line of centers, and $A$ and $B$ relate the relative velocity to the rate of
strain. Their expressions in terms of the two-sphere mobility and resistance
functions are given in Appendix~\ref{app:numerical-method}.
The imposed velocity gradient and rate of strain are normalized by
the shear rate, $\hat{\boldsymbol{\Gamma}} = \langle\boldsymbol{\Gamma}\rangle/\dot\gamma$
and $\hat{\mathbf{E}} = \langle\mathbf{E}\rangle/\dot\gamma$, and
$\hat{\boldsymbol{\Omega}} = \hat{\boldsymbol{\Gamma}} - \hat{\mathbf{E}}$ is
the rotational part. The second form of Eq.~\eqref{eq:Urel} shows that $1-A$
and $1-B$ scale the relative velocity driven by the strain along and
transverse to the line of centers, respectively.

In the weak-flow limit the distortion is linear in $\mathrm{Pe}_b$, see
Refs.~\onlinecite{batchelorEffectBrownianMotion1977,russelColloidalDispersions1989},
\begin{equation}
  g(\mathbf{r}) = g_{\mathrm{eq}}(r)\left[\,1
    - \mathrm{Pe}_b\, f(\rho)\,\hat{\mathbf{r}}\cdot\hat{\mathbf{E}}\cdot\hat{\mathbf{r}}
    + \mathcal{O}(\mathrm{Pe}_b^{2})\right],
  \label{eq:distortion}
\end{equation}
with $g_{\mathrm{eq}}=\Theta(r-2b)$ the dilute equilibrium pair distribution
of Eq.~\eqref{eq:pair_eq}.
Equation~\eqref{eq:smol} reduces to a linear two-point boundary-value problem for the
scalar distortion~\cite{bergenholtzNonNewtonianRheologyDilute2002,batchelorEffectBrownianMotion1977, russelHugginsCoefficientMeans1984}
\begin{equation}
  \frac{1}{\rho^{2}}\frac{\mathrm{d}}{\mathrm{d}\rho}
    \!\left[\rho^{2}G(\Lambda\rho)\,\frac{\mathrm{d}f}{\mathrm{d}\rho}\right]
  - \frac{6H(\Lambda\rho)}{\rho^{2}}\,f(\rho)
  = -W(\Lambda\rho),
  \label{eq:bvp}
\end{equation}
subject to
\begin{equation}
  G(2 \Lambda)\left.\frac{\mathrm{d}f}{\mathrm{d}\rho}\right|_{\rho=2}
    = -2\left[1 - A(2 \Lambda)\right],
  \qquad f \to 0 \quad (\rho\to\infty),
  \label{eq:bvpbc}
\end{equation}
where $W(s) \equiv -3\left[A(s) - B(s)\right] - s\,A'(s)$.

Substituting Eq.~\eqref{eq:distortion} into the pair-level forms of
Eqs.~\eqref{eq:SB_def} and \eqref{eq:SP_def} and evaluating their deviatoric
parts~\cite{batchelorEffectBrownianMotion1977,russelHugginsCoefficientMeans1984,bradyRheologicalBehaviorConcentrated1993,bergenholtzNonNewtonianRheologyDilute2002}
gives
$\mathcal{A} = \mathcal{A}_P + \mathcal{A}_B$ with
%
%
\begin{align}
  \mathcal{A}_P(\Lambda)
    &= \frac{9}{5\Lambda}\left[1 - A(2 \Lambda)\right] f(2),
  \label{eq:AP}\\[2pt]
  \mathcal{A}_B(\Lambda)
    &= \frac{9}{40\Lambda}\int_{2}^{\infty}
       W(\Lambda\rho)\, f(\rho)\, \rho^{2}\,\mathrm{d}\rho .
  \label{eq:AB}
\end{align}
The derivation is given in Appendix~\ref{app:pair-stress}.
For $\Lambda>1$ both contributions are positive, since $f > 0$, $W > 0$, and $1-A(2\Lambda) > 0$.
At $\Lambda=1$ the interparticle contribution vanishes, because $A(2)=1$.
The numerical solution of Eqs.~\eqref{eq:bvp} and \eqref{eq:bvpbc} and the
evaluation of Eqs.~\eqref{eq:AP} and \eqref{eq:AB} are described in
Appendix~\ref{app:numerical-method}, and a closed-form large-$\Lambda$
expansion is given in Appendix~\ref{app:large-Lambda}.
Equation~\eqref{eq:Adef} is a pair result and holds only to $\mathcal{O}(\phi_b^{2})$.
Extending it requires a closure.
Following Ref.~\onlinecite{russelDivergenceLowShear2013a}, we use two closures that place the divergence of $\eta_T$ at different volume fractions, the glass transition and random close packing.
Both write $\eta_T$ as a pair amplitude times a divergence factor, so that the exact $\mathcal{A}(\Lambda)$ can be inserted directly.

\subsubsection*{The MCT closure}

We now turn to \gls{mct} to extend Eq.~\eqref{eq:Adef} to higher
concentrations. \Gls{mct} describes structural relaxation through the decay of
coupled density fluctuations. It predicts a bifurcation at an ideal glass
transition $\phi_g$. This marks the point where the $\alpha$-relaxation time
diverges and long-time diffusion vanishes, while the microscopic short-time
dynamics remains finite and nonsingular~\cite{gotzeRelaxationProcessesSupercooled1992}.
The $\alpha$-relaxation time for the one-component hard-sphere \gls{mct} model with
a Percus--Yevick static structure factor~\cite{franoschAsymptoticLawsPreasymptotic1997} diverges with exponent
$\gamma=2.46$. Since the
thermodynamic viscosity is governed by the same diverging structural-relaxation
time scale, it has the corresponding asymptotic concentration dependence~\cite{russelDivergenceLowShear2013a}
\begin{equation}
  \frac{\eta_{T}}{\mu} \sim
   \left(1 - \frac{\phi_b}{\phi_g}\right)^{-2.46},
   \qquad
   \phi_b \rightarrow \phi_g.
  \label{eq:mc_div}
\end{equation}
Combining this diverging behavior with the low-volume-fraction expression
of Eq.~\eqref{eq:Adef}, we may write~\cite{russelDivergenceLowShear2013a}
\begin{equation}
    \frac{\eta_{T}}{\mu} =
    \mathcal{A}(\Lambda) \phi_b^2
    \left(1 - \frac{\phi_b}{\phi_g}\right)^{-2.46}.
    \label{eq:mct}
\end{equation}
With structure factors from the Percus--Yevick closure the
theory~\cite{franoschAsymptoticLawsPreasymptotic1997} places
$\phi_g \simeq 0.52$.
In practice $\phi_g$ is rescaled to $\simeq 0.57$ to
absorb the known approximations and to match the arrest observed by dynamic
light scattering~\cite{vanmegenArrestFlowEmergence2009,russelDivergenceLowShear2013a}.

This closure contains no hydrodynamics apart from the amplitude $\mathcal{A}(\Lambda)$.
Its divergence is purely structural.
It occurs below random close packing, and its exponent is independent of $\Lambda$.
In Ref.~\onlinecite{russelDivergenceLowShear2013a} the amplitude is $0.9$,
close to the exact hard-sphere value $\mathcal{A}(1) = 0.913$ but independent of $\Lambda$.
Replacing it by $\mathcal{A}(\Lambda)$ makes the dilute limit exact at every
$\Lambda$.

\subsubsection*{The Brady closure}

Brady~\cite{bradyRheologicalBehaviorConcentrated1993} showed that, in weak flow, the thermodynamic stress is proportional to the equilibrium pair distribution at contact and to the departure of the pair distribution from equilibrium.
This departure scales with a P\'eclet number based on the short-time self-diffusivity, which includes the hydrodynamic interactions with the surrounding particles.
The thermodynamic viscosity therefore scales as the contact value divided by the short-time self-diffusivity.
For the present
two-radius model, the equilibrium structure is set by the thermodynamic
volume fraction $\phi_b$, whereas the short-time self-diffusivity is evaluated
at the hydrodynamic volume fraction $\phi_a$. With the equilibrium contact value
\begin{equation}
  g_c(\phi_b) \equiv g_{\mathrm{eq}}(r=2b;\phi_b),
\end{equation}
Brady's excluded-shell result~\cite{bradyRheologicalBehaviorConcentrated1993} can be written as
\begin{equation}
  \frac{\eta_T}{\mu}
  =
  \frac{12}{5\Lambda}\,
  \phi_b^2
  \frac{g_c(\phi_b)}
       {\hat D^{\mathrm{s}}(\phi_a;\Lambda)} ,
  \label{eq:brady}
\end{equation}
where $\hat D^{\mathrm{s}}=D^{\mathrm{s}}/D_0$ is the short-time self-diffusivity reduced by the
single-particle diffusivity $D_0$ of Eq.~\eqref{eq:Peb}.
In obtaining this expression, the short-time self-diffusivity is
first scaled out of the pair dynamics and the
remaining pair hydrodynamic interactions are then
neglected. The prefactor $12/(5\Lambda)$ is consequently a
free-draining result.

The contact value follows the Carnahan--Starling equation of
state~\cite{carnahanEquationStateNonattracting1969} at lower concentrations
and a close-packing asymptote at higher concentrations.
We place the close-packing pole at $\phi_m=\phi_{\mathrm{div}}(1)=[\tfrac{15}{2}I(1)]^{-1/2}\simeq0.632$, so that the contact and hydrodynamic poles coincide exactly at $\Lambda=1$.
Brady~\cite{bradyRheologicalBehaviorConcentrated1993} switched between the two branches at $\phi=0.5$.
We instead join them where they cross, following the continuity argument of Torquato~\cite{torquatoNearestneighborStatisticsPackings1995}, but keep Brady's prefactor $1.2$,
\begin{equation}
  g_c(\phi_b)=
  \begin{cases}
    \displaystyle
    \frac{1-\phi_b/2}{(1-\phi_b)^3}, & \phi_b<\phi_*, \\[8pt]
    \displaystyle
    \frac{1.2}{1-\phi_b/\phi_m}, & \phi_*\le\phi_b<\phi_m,
  \end{cases}
  \label{eq:pc}
\end{equation}
where continuity gives $\phi_*\simeq0.517$.
Rather than using simulation values of the short-time self-diffusivity, as
in Ref.~\onlinecite{bradyRheologicalBehaviorConcentrated1993},
we evaluate $\hat D^{\mathrm{s}}$ from the effective-medium
relation of Lionberger and Russel~\cite{lionbergerMicroscopicTheoriesRheology1999}
\begin{equation}
  \hat D^{\mathrm{s}}(\phi_a;\Lambda)
  =
  \frac{3}{1+2\eta'_\infty(\phi_a;\Lambda)/\mu},
  \label{eq:diffusivity_eta_infty}
\end{equation}
with $\eta'_\infty$ taken from Eq.~\eqref{eq:eta_inf}. Through $I(\Lambda)$,
$\hat D^{\mathrm{s}}$ therefore depends on $\Lambda$ as well as on $\phi_a$,
and vanishes at the same $\phi_a$ as the divergence of $\eta'_\infty$.

The free-draining prefactor in Eq.~\eqref{eq:brady} is
not generally the correct pair amplitude.
Indeed, setting $G=H=1$ and $A=B=0$ in
Eqs.~\eqref{eq:bvp}--\eqref{eq:AB} gives
$\mathcal A=12/(5\Lambda)$, whereas retaining the full two-particle
hydrodynamics gives $\mathcal A(\Lambda)=\mathcal A_P+\mathcal A_B$.
At $\Lambda=1$, for example, the former is $2.4$, compared with the
exact pair value $\mathcal A(1)=0.913$.

We therefore retain the concentration dependence of Brady's closure but
replace its free-draining pair amplitude by the full dilute result,
\begin{equation}
  \frac{\eta_T}{\mu}
  =
  \mathcal A(\Lambda)\,
  \phi_b^2
  \frac{g_c(\phi_b)}
       {\hat D^{\mathrm{s}}(\phi_a;\Lambda)},
  \qquad
  \phi_a=\Lambda^{-3}\phi_b .
  \label{eq:closure_ours}
\end{equation}
Because $g_c\to1$ and $\hat D^{\mathrm{s}}\to1$ as $\phi_b\to0$,
Eq.~\eqref{eq:closure_ours} recovers the exact $\mathcal{O}(\phi_b^2)$ pair
result for every $\Lambda$.

\subsection{Hard-sphere simulations}
\label{subsec:hard_sphere_simulations}

We validate the hydrodynamic contribution of the two-radius model,
Eq.~\eqref{eq:eta_inf}, against numerical simulations. This
comparison is particularly useful because experimental determinations of
the high-frequency viscosity of hard-sphere suspensions are relatively
scarce. Unlike the zero-shear viscosity, the high-frequency limit is
generally not directly accessible within the frequency range of
conventional bulk rheometry. It must therefore be inferred indirectly
by fitting the complex modulus over an accessible frequency
range~\cite{vanderwerffLinearViscoelasticBehavior1989}, or accessed using
microrheological techniques~\cite{laermannEmergentSignaturesGlass2026}.

Because the high-frequency viscosity is an average over the equilibrium
microstructure, its calculation requires accurate sampling of the
hard-sphere equilibrium distribution, particularly at small interparticle
separations where near-field hydrodynamic interactions are strongest.
Several finite-time-step \gls{bd} schemes for hard spheres have been
proposed~\cite{cichockiDynamicComputerSimulation1990,schaertlBrownianDynamicsSimulations1994,heyesBrownianDynamicsSimulations1993},
but some have been found to bias the sampled equilibrium
structure~\cite{stratingBrownianDynamicsSimulation1999}.
We use the method of Strating~\cite{stratingBrownianDynamicsSimulation1999}, in which overlaps are prevented by elastic binary collisions within each time step, so that the hard-sphere constraint is enforced exactly.
All simulations were performed using custom
implementations developed within the \texttt{JAX-MD}
framework~\cite{schoenholzJAXMDFramework2021}.

The equilibrium distribution of the particle
coordinates reduced by $b$ depends only on $\phi_b$ and is therefore
independent of $\Lambda$. We consequently generated a single ensemble of
equilibrium configurations for each $\phi_b$, using $b=a$, and reused the same reduced
coordinates for all five values of $\Lambda$ ($1$, $1.02$, $1.06$, $1.1$ and $1.2$). The resulting systems therefore
have identical thermodynamic microstructures, while differing in the
minimum separation between hydrodynamic surfaces.

For each $\phi_b$, twenty statistically independent \gls{bd}
trajectories were initialized from uniformly distributed random particle
coordinates in a cubic periodic box. Initial overlaps were removed by
minimizing a soft-sphere penalty potential using the \gls{fire}
algorithm~\cite{bitzekStructuralRelaxationMade2006}.
Each trajectory, with $N=512$ particles, was then run for $50\,b^2/D_0$ with a time step of $0.005\,b^2/D_0$, and its final configuration was used, giving twenty configurations per $\phi_b$.
Hydrodynamic
interactions were omitted during this equilibration stage because
they modify the dynamics of relaxation but not the
equilibrium configurational distribution.

For each sampled configuration, the high-frequency viscosity was evaluated
directly from the hydrodynamic stress (Eq.~\eqref{eq:SH_def}). Hydrodynamic interactions and
particle stresslets were computed using the \gls{fsd}
formulation~\cite{fioreFastStokesianDynamics2019}, which we implemented in
\texttt{JAX-MD}, drawing on the implementation of the \texttt{jfsd}
package~\cite{torrePythonJAXbasedFastStokesian2025}. This follows the standard
\gls{sd} procedure for calculating suspension transport
properties and the high-frequency viscosity~\cite{bradyStokesianDynamics1988,phillipsHydrodynamicTransportProperties1988,fossStructureDiffusionRheology2000}.

%% file: sections/results_and_discussion.tex
\section{Results and discussion}
\label{sec:results_discussion}
We first calculate the hydrodynamic part of the zero-shear viscosity, from the
pair kernel to finite concentrations, and compare it with our simulations. We
then calculate the dilute amplitude of the thermodynamic part and compare it
with the hydrodynamic pair coefficient. Next, we use the two closures to obtain
the thermodynamic part at finite concentrations and combine both parts to get
the zero-shear viscosity. Finally, we compare the two-radius model with
measurements on experimental suspensions of sterically and charge-stabilized particles.

\subsection{Hydrodynamic viscosity}
\label{subsec:res_hydrodynamic}

In this section, we evaluate the high-frequency viscosity from Eq.~\eqref{eq:eta_inf}. The
excluded shell enters only through the lower limit $s=2\Lambda$ of the integral
$I(\Lambda)$, so we first discuss the kernel $J(s)$ before covering its effect on the integral.
Figure~\ref{fig:I_of_x_a} shows the kernel as a function of the hydrodynamic surface
gap $\xi \equiv s-2$. It assumes its maximum value at contact and decays monotonically over about one
hydrodynamic radius. The far-field form~\cite{batchelorDeterminationBulkStress1972}
$J\simeq\tfrac{15}{2}s^{-6}$ follows $J$ closely for $\xi\gtrsim0.5$,
but it cannot capture the near-field behavior and the contact value.
Batchelor and Green~\cite{batchelorDeterminationBulkStress1972} interpolated
between the contact value and the far field. Yoon and
Kim~\cite{yoonNoteDirectCalculation1987} later showed that $J$ approaches $J(2)$
only logarithmically, and fitted a series in $1/\ln(1/\xi)$ to numerical data.
Inverting the lubrication forms of the resistance
functions~\cite{jeffreyCalculationResistanceMobility1984,
townsendGeneratingScratchNearfield2023}, we find the exact asymptotic form as
$\xi \to 0$ to be
\begin{equation}
  J = 0.2214 - \frac{0.3002}{\ln(1/\xi)+1.347} + \mathcal{O}(\xi\ln\xi).
  \label{eq:J_contact}
\end{equation}
Equation~\eqref{eq:J_contact} follows $J$ closely over many decades of the gap $\xi$ and
departs from it only for $\xi \gtrsim 10^{-2}$. This leaves the region $10^{-2} \lesssim \xi \lesssim 0.5$ where $J$ needs to be determined numerically.

\begin{figure}[!tb]
  \panel{fig:I_of_x_a}{%
    \includegraphics[scale=1]{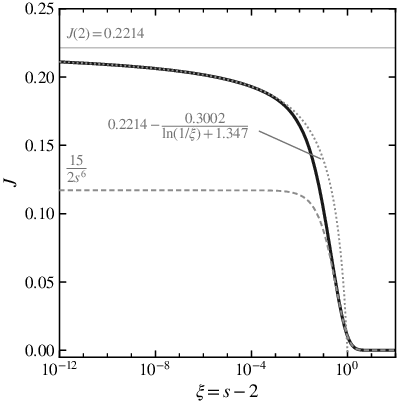}}\hfill
  \panel{fig:I_of_x_b}{%
    \includegraphics[scale=1]{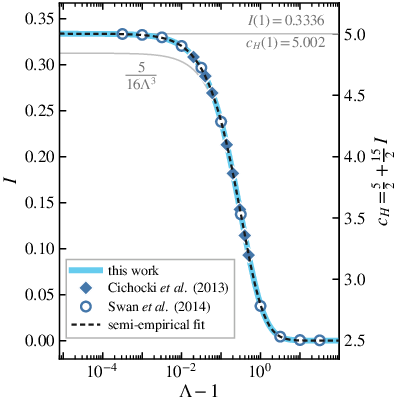}}
  \caption{Pair-level functions for computing the high-frequency viscosity.
  (a) The stresslet kernel~\cite{batchelorDeterminationBulkStress1972} $J$,  Eq.~\eqref{eq:J_mobility},
  against the hydrodynamic surface gap $\xi=s-2$. The
  gray lines and curves show the contact value $J(2) \approx 0.2214$ (solid), the near-contact form of
  Eq.~\eqref{eq:J_contact} (dotted), and the far-field form $J \approx 15/(2s^{6})$ (dashed).
  (b) The pair integral $I$, Eq.~\eqref{eq:eta_inf}, as a function of  the excluded shell
  gap $\Lambda-1$. The right axis shows the corresponding values of $c_H$,
  Eq.~\eqref{eq:BG_c2}.
  Filled symbols show the annulus-model values of $c_H$ by Cichocki \textit{et al.}%
  ~\cite{cichockiShorttimeDynamicsHighfrequency2013}, and open circles the
  high-frequency Huggins coefficient, $\kappa_\infty$, digitized from Swan \textit{et al.}%
  ~\cite{swanMediumAmplitudeOscillatory2014}, $c_H=\tfrac{5}{2}\kappa_\infty$. The black
  dashed curve is the semi-empirical fit, Eq.~\eqref{eq:I_approx}. The gray line and curve provide
  the hard-sphere value, $I(1) \approx 0.3336$ and $c_H(1) \approx 5.002$, and the large-$\Lambda$
  asymptotic expression $I \approx 5/(16 \Lambda^3)$, Eq.~\eqref{eq:I_large_Lambda}, respectively.}
  \label{fig:I_of_x}
\end{figure}

Figure~\ref{fig:I_of_x_b} shows $I$ and $c_H$ as functions of the
excluded shell gap $\Lambda-1$. The results agree with the hard-sphere value~\cite{cichockiShorttimeDiffusionCoefficients1988,
cichockiShorttimeDynamicsHighfrequency2013}
$c_H(1)=5.002$ and reproduce the annulus-model values of Cichocki
\textit{et al.}~\cite{cichockiShorttimeDynamicsHighfrequency2013} and of Swan
\textit{et al.}~\cite{swanMediumAmplitudeOscillatory2014} over the whole range
of $\Lambda$.
The integral $I$ decreases monotonically with $\Lambda - 1$. Increasing the size of the excluded shell removes the configurations
closest to contact, where $J$ is largest, so even a thin shell lowers $I$ appreciably.
For large $\Lambda$ only the far field of the kernel contributes, leading to the scaling
\begin{equation}
  I(\Lambda)\simeq\frac{5}{16\Lambda^{3}}.
  \label{eq:I_large_Lambda}
\end{equation}
This far-field approximation was used by Russel~\cite{russelHugginsCoefficientMeans1984}
for the hydrodynamic stress of the excluded-shell model at every $\Lambda$.
Since the far-field form describes $J$ closely outside the lubrication region,
this approximation is accurate for thick shells (Fig.~\ref{fig:I_of_x_b}), and
even near contact it underestimates $I$ only slightly.
To cover the whole range we introduce a new semi-empirical expression
\begin{equation}
  I(\Lambda) \approx \frac{5}{16\Lambda^{3}}
  \left[1+\frac{p_0}{1+q_1\zeta+q_2\zeta^{2}}\right],
  \qquad \zeta=\Lambda-1.
  \label{eq:I_approx}
\end{equation}
This form recovers the far-field result, Eq.~\eqref{eq:I_large_Lambda}, as $\Lambda\to\infty$.
Choosing $p_0 = 16I(1)/5-1 \simeq 0.0676$ reproduces the hard-sphere value $I(1)$. The
remaining two constants follow from a least-squares fit to the exact integral,
which gives $q_1=20$ and $q_2=127$.
The resulting curve (dashed in Fig.~\ref{fig:I_of_x_b}) follows the exact integral
$I$ closely from contact to the far-field limit.

\begin{figure*}[tb]
  \centering
  \includegraphics[scale=1]{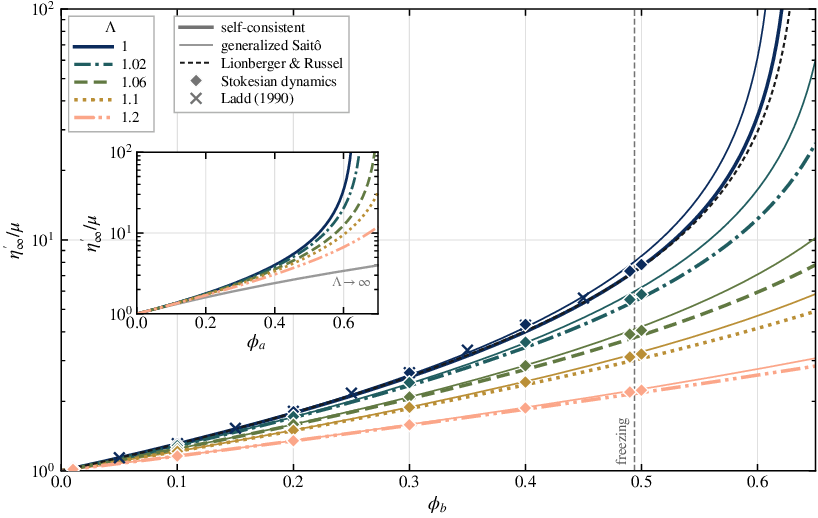}
  \caption{High-frequency viscosity $\eta'_{\infty}$ against the thermodynamic volume fraction
  $\phi_b=\Lambda^{3}\phi_a$, for several values of $\Lambda$.
  Thick curves show the outcome of applying the self-consistent closure, Eq.~\eqref{eq:eta_inf},
  and thin curves in the same color represent the outcome of the generalized
  Saitô formula, Eq.~\eqref{eq:saito}. The black dashed curve follows by applying the formula of Lionberger and
  Russel~\cite{lionbergerHighFrequencyModulus1994} for pure hard spheres ($\Lambda=1$).
  Diamonds show our \gls{sd} results for every value of
  $\Lambda$ in the corresponding color (the numerical uncertainty is smaller than the
  symbol size), and the crosses show the simulation results obtained by Ladd~\cite{laddHydrodynamicTransportCoefficients1990}
  for $\Lambda=1$. The dashed vertical line marks the hard-sphere freezing volume
  fraction~\cite{puseyHardSpheresCrystallization2009}, $\phi_b=0.494$.
  The inset shows Eq.~\eqref{eq:eta_inf} against the hydrodynamic volume fraction
  $\phi_a$, together with its limit $\Lambda\to\infty$.}
  \label{fig:eta_inf_theory}
\end{figure*}

Figure~\ref{fig:eta_inf_theory} shows the high-frequency viscosity predicted
by the self-consistent closure against the thermodynamic volume fraction
$\phi_b$. The equilibrium microstructure depends only on $\phi_b$, so at fixed $\phi_b$ all
values of $\Lambda$ describe the same structure, with the same critical
volume fractions, and differ only in their hydrodynamics.
In this representation, Eq.~\eqref{eq:eta_inf} becomes
\begin{equation}
  \frac{\eta'_\infty}{\mu}
  = \frac{1 + \tfrac{5}{2}\,\Lambda^{-3}\phi_b + \tfrac{5}{2}\,\Lambda^{-6}\phi_b^{2}}
         {1 - \tfrac{15}{2}\,\Lambda^{-6}I(\Lambda)\,\phi_b^{2}},
  \label{eq:eta_inf_phi_b}
\end{equation}
so the shell enters both through the powers of $\Lambda^{-3}$, which decrease the contributions of
the hydrodynamic interactions, and through $I$. The inset shows the same
curves against $\phi_a$, which isolates the effect of
$I$. The curves separate increasingly as the volume fraction grows.
At $\Lambda=1$, Eq.~\eqref{eq:eta_inf} reduces to the closure of
Potanin~\cite{potaninSelfConsistentCalculationsViscosity1993}, who used the
pair coefficient $5.2$ as calculated by Batchelor and Green%
~\cite{batchelorDeterminationBulkStress1972}. This positioned the pole of $\eta'_\infty$ 
at $\phi_b=0.61$, which is below the random close packing fraction of hard
spheres~\cite{scottDensityRandomClose1969,berrymanRandomClosePacking1983,torquatoJammedHardparticlePackings2010}.
The exact coefficient $5.002$ moves it to $0.632$, close to random close
packing. This is where
$\eta'_\infty$ is expected to diverge, as the gaps between neighboring particles
close and the lubrication resistances become singular%
~\cite{bradyRheologicalBehaviorConcentrated1993}. The closure obtains this
divergence from pair hydrodynamics alone, with many-body effects included only
through a mean-field approximation.
For $\Lambda>1$, $\eta'_\infty$ remains finite at random close packing, because touching hard excluded shells still leave a gap of $2(b-a)$ between the hydrodynamic surfaces.

Figure~\ref{fig:eta_inf_theory} includes, for comparison, the generalized
Saitô
formula~\cite{abadeHighfrequencyViscosityGeneralized2010,riestDynamicsSuspensionsHydrodynamically2015},
which extends Saitô's mean-field
result~\cite{saitoConcentrationDependenceViscosity1950},
$(\eta'_\infty-\mu)/(\eta'_\infty+\tfrac32\mu)=\phi_a$, using a correction factor $S$
which is fixed using the exact pair coefficient. In the present representation, the following high-frequency viscosity is obtained
\begin{equation}
  \frac{\eta'_\infty}{\mu}
  = 1+\frac{5}{2}\,\frac{\phi_a(1+S)}{1-\phi_a(1+S)},
  \qquad
  S=3I(\Lambda)\,\phi_a .
  \label{eq:saito}
\end{equation}
Like Eq.~\eqref{eq:eta_inf}, it reproduces $c_H$ by construction,
so the two closures differ only beyond $\mathcal{O}(\phi_a^{2})$. Equation~\eqref{eq:saito} lies above
Eq.~\eqref{eq:eta_inf} and diverges earlier. For pure hard spheres, Lionberger
and Russel~\cite{lionbergerHighFrequencyModulus1994} used the same general form with
the empirical correction $S=\phi_a-0.189\phi_a^{2}$.

Lastly, Fig.~\ref{fig:eta_inf_theory} shows $\eta'_\infty$ as obtained using
our \gls{sd} simulations for every value of $\Lambda$. These numerical results agree
with the simulations of Ladd~\cite{laddHydrodynamicTransportCoefficients1990} for $\Lambda = 1$.
For every $\Lambda$, the simulations follow the self-consistent closure, Eq.~\eqref{eq:eta_inf}, up to freezing.
The simulations are limited to $\phi_b\le0.50$, because at higher volume fractions the system crystallizes during equilibration. This crystallinity interferes with the comparison to the theoretical values, which assume that the system remains amorphous. Polydispersity is required to suppress the tendency of the system to crystallize~\cite{puseyHardSpheresCrystallization2009} and can be used to render it glassy~\cite{zaccarelliPolydispersityHardSphere2015}. However, in that case the theory may also need to be modified to account for polydispersity explicitly, which we leave for future work.

\subsection{Dilute amplitude of the thermodynamic viscosity}
\label{subsec:res_amplitude}

\begin{figure*}[tb]
  \centering
  \includegraphics[scale=1]{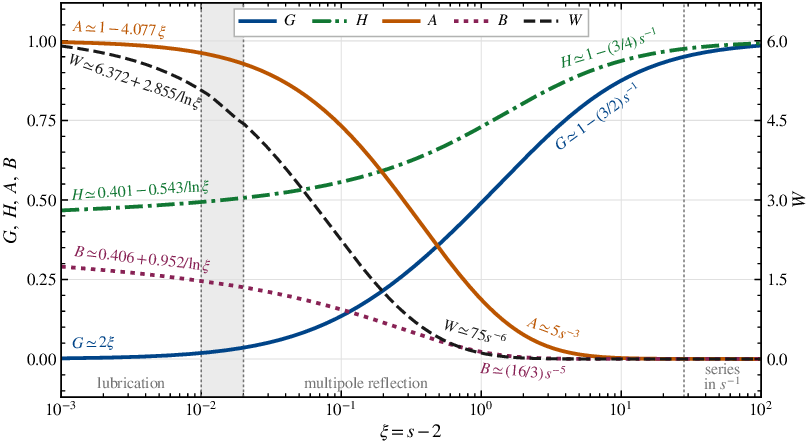}
  \caption{Hydrodynamic functions of the pair Smoluchowski problem plotted
  against the dimensionless hydrodynamic surface gap $\xi \equiv s-2$.
  Dotted lines mark where the numerical calculation changes method
  (Appendix~\ref{app:numerical-method}), as labeled in gray along the $\xi$-axis, and the shaded band marks where the
  two methods are joined by interpolation. All five functions are continuous across the band, and only $W$ shows a slight change of slope there. The leading lubrication forms,
  derived from the resistance functions~\cite{
    jeffreyCalculationResistanceMobility1984,jeffreyCalculationLowReynolds1992,
    townsendGeneratingScratchNearfield2023},
  are written along the curves at the left and the leading terms of the
  far-field analytic
  approximations~\cite{batchelorBrownianDiffusionParticles1976,batchelorEffectBrownianMotion1977}
  at the right.}
  \label{fig:fig_pair_functions}
\end{figure*}

Now that we have covered the high-frequency viscosity, we turn to the thermodynamic viscosity, $\eta_T$. We start by computing the dilute thermodynamic viscosity amplitude
$\mathcal{A}(\Lambda)$ by solving Eqs.~\eqref{eq:bvp}--\eqref{eq:AB}.
The hydrodynamic functions required to obtain $\mathcal{A}$ are $A$, $B$, $G$, $H$, and $W$, which are shown in
Fig.~\ref{fig:fig_pair_functions} together with their lubrication forms
near contact and their analytic approximation for large separations. Since
the excluded shells keep the particle centers at least $s=2\Lambda$ apart,
only the part of these functions with $\xi=s-2\geq2(\Lambda-1)$ enters
$\mathcal{A}$.
Figure~\ref{fig:A_of_x_a} shows the value of the distortion at the excluded shell,
$f(2)$, which enters the interparticle contribution through Eq.~\eqref{eq:AP}.
Figure~\ref{fig:A_of_x_b} shows the amplitude $\mathcal{A}$ and its
interparticle and Brownian contributions as functions
of the shell gap $\Lambda-1$. The results agree with the values of Bergenholtz
\textit{et al.}~\cite{bergenholtzNonNewtonianRheologyDilute2002} and with the two 
asymptotic limits, $\mathcal{A}(1)=0.913$ for pure hard spheres%
~\cite{cichockiLongtimeSelfdiffusionCoefficient1988} and
$\mathcal{A}(\Lambda)\sim12/(5\Lambda)$ for widely separated spheres%
~\cite{russelHugginsCoefficientMeans1984,bradyRheologicalBehaviorConcentrated1993}.
Swan \textit{et al.}~\cite{swanMediumAmplitudeOscillatory2014} reported the
low- and high-frequency Huggins coefficients of the same model, $\kappa_0$ and
$\kappa_\infty$, defined on the hydrodynamic volume fraction. Their difference
gives the thermodynamic amplitude,
$\mathcal{A}=\tfrac{5}{2}(\kappa_0-\kappa_\infty)\Lambda^{-6}$, which is shown
in Fig.~\ref{fig:A_of_x_b}. It agrees with ours for $\Lambda$ close to 1 and
for large $\Lambda$, and lies slightly above it around the maximum. Their
$\kappa_\infty$ follows our $c_H$ over the whole range
(Fig.~\ref{fig:I_of_x_b}), so the difference arises in the solution of the
pair Smoluchowski problem, Eqs.~\eqref{eq:bvp}--\eqref{eq:AB}. A possible
origin is the hydrodynamic functions at intermediate separations, which Swan
\textit{et al.} calculated from the expressions of Jeffrey and
Onishi~\cite{jeffreyCalculationResistanceMobility1984}. The recurrence
relations printed there contain misprints, which were corrected
only recently~\cite{townsendGeneratingScratchNearfield2023}.
Away from contact, our functions are computed with the multipole-reflection
method~\cite{wilsonStokesFlowThree2013} and do not rely on these series.

\begin{figure}[!tb]
  \panel{fig:A_of_x_a}{%
    \includegraphics[scale=1]{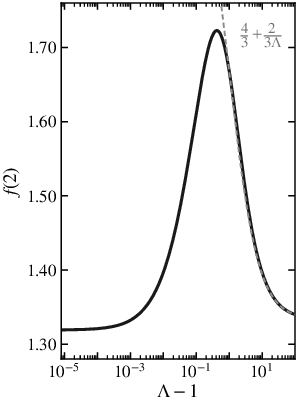}}\hfill
  \panel{fig:A_of_x_b}{%
    \includegraphics[scale=1]{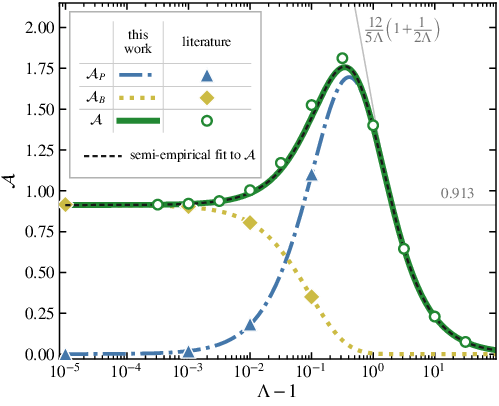}}
  \caption{Results from the numerical solution of Eqs.~\eqref{eq:bvp}--\eqref{eq:AB},
  plotted against the excluded shell gap $\Lambda-1$.
  (a) Value of the distortion at the excluded shell, $f(2)$. The
  dashed curve shows the first two terms of its large-$\Lambda$ approximation
  (Appendix~\ref{app:large-Lambda}).
  (b) Dilute amplitude of the thermodynamic viscosity, $\mathcal{A}$ (green,
  solid), and its interparticle, $\mathcal{A}_P$ (blue, dash-dotted), and
  Brownian, $\mathcal{A}_B$ (yellow, dotted), contributions. Filled triangles
  ($\mathcal{A}_P$) and diamonds ($\mathcal{A}_B$) show values digitized from
  Bergenholtz \textit{et al.}~\cite{bergenholtzNonNewtonianRheologyDilute2002},
  and open circles the amplitude $\mathcal{A}$ obtained from the Huggins coefficients
  digitized from Swan \textit{et al.}~\cite{swanMediumAmplitudeOscillatory2014}.
  The black dashed curve is the
  semi-empirical fit, Eq.~\eqref{eq:A_approx}. The two asymptotes are shown as
  gray guides to the eye.}
  \label{fig:A_of_x}
\end{figure}

In the hard-sphere limit, $\Lambda\to1^{+}$, the amplitude is entirely Brownian,
$\mathcal{A}(1)=\mathcal{A}_B(1)\simeq0.913$, and the interparticle contribution vanishes.
Lubrication forces bring the relative velocity of two hard spheres along
their centerline to zero as they
touch, so the flow cannot push them into contact in a finite
time~\cite{batchelorHydrodynamicInteractionTwo1972,russelColloidalDispersions1989}.
The hard-sphere force therefore causes no relative motion, and it adds no
stress~\cite{batchelorEffectBrownianMotion1977}.
This can be seen directly in Eq.~\eqref{eq:AP}, where the interparticle
contribution is proportional to $1-A(2\Lambda)$, which by
Eq.~\eqref{eq:Urel} measures that relative velocity and vanishes at contact,
$A(2)=1$.

For $\Lambda-1\lesssim10^{-3}$ neither contribution changes appreciably,
because the excluded shell opens only a very thin hydrodynamic gap,
$\xi=2(\Lambda-1)$, between touching spheres.
The interparticle contribution grows
with $1-A(2\Lambda) \propto (\Lambda-1)$, since $f(2)$ is almost
constant (Fig.~\ref{fig:A_of_x_a}). The Brownian contribution loses only the thin
slice of separations next to contact, across which $W$
and $f$ vary little. Both changes are therefore approximately
linear in the small gap $\Lambda-1$.

For thicker shells both contributions change appreciably, and the amplitude
becomes nonmonotonic. The excluded shell now holds
pairs outside the lubrication region, and the factor $1-A(2\Lambda)$ grows
quickly toward $1$ (Fig.~\ref{fig:fig_pair_functions}), while $f(2)$ changes
comparatively little. This results in a steep rise of the interparticle
contribution. The Brownian contribution falls, because the separations
excluded by the shell are those where $W$ is largest.
The amplitude peaks near $\Lambda=1.35$,
at nearly twice the pure hard-sphere value.

Beyond the peak, the growth of $1-A(2\Lambda)$ no longer outweighs
the prefactor $1/\Lambda$ in Eq.~\eqref{eq:AP}, and the amplitude goes
toward the free-draining limit~\cite{russelHugginsCoefficientMeans1984,bradyRheologicalBehaviorConcentrated1993}
$12/(5\Lambda)$.
Expanding the pair problem in powers of $\Lambda^{-1}$ gives
(Appendix~\ref{app:large-Lambda})
\begin{equation}
  \mathcal{A}(\Lambda)
  = \frac{12}{5\Lambda}\left[1 + \frac{1}{2\Lambda}
    + \frac{31}{112\,\Lambda^{2}} - \frac{4003}{3584\,\Lambda^{3}}
    + \mathcal{O}\!\left(\Lambda^{-4}\right)\right],
  \qquad
  \mathcal{A}_B(\Lambda) \simeq \frac{15}{32}\,\Lambda^{-7}.
  \label{eq:A_large_Lambda}
\end{equation}
The first correction is positive, showing that hydrodynamic interactions increase
$\mathcal{A}$ at large $\Lambda$. The Brownian contribution, in
contrast, decays as $\Lambda^{-7}$, and at large $\Lambda$ the amplitude is
purely interparticle.

To use $\mathcal{A}$ in the closures of Eqs.~\eqref{eq:mct} and
\eqref{eq:closure_ours} without solving the pair problem for every $\Lambda$, we
approximate it by a new semi-empirical function of the shell gap $\zeta\equiv\Lambda-1$,
\begin{equation}
  \mathcal{A}(\Lambda) \approx \frac{12}{5\Lambda}\,
  \frac{a_0 + a_1\zeta + a_2\zeta^{2} + \zeta^{3}}
       {1 + b_1\zeta + b_2\zeta^{2} + \zeta^{3}}.
  \label{eq:A_approx}
\end{equation}
This form is chosen because it can reproduce both limits exactly. Equal
leading $\zeta^3$ coefficients give the free-draining limit, and the choices
$a_2=b_2+\tfrac12$ and $a_1=b_1+\tfrac12b_2-\tfrac{25}{112}$ reproduce the
$\Lambda^{-1}$ and $\Lambda^{-2}$ terms in the bracket of
Eq.~\eqref{eq:A_large_Lambda}. The constant
$a_0=5\mathcal{A}(1)/12\simeq0.380$ reproduces the hard-sphere value $\mathcal{A}(1)$. Only $b_1$ and
$b_2$ are then free. We choose them to minimize the largest relative deviation
from the numerical solution. This gives $b_1=3.4$ and $b_2=3.5$, and therefore
$a_1=4.927$ and $a_2=4$. The resulting curve (dashed in
Fig.~\ref{fig:A_of_x_b}) follows the numerical solution closely from contact to
the free-draining limit.

\begin{figure}[!tb]
  \includegraphics[scale=1]{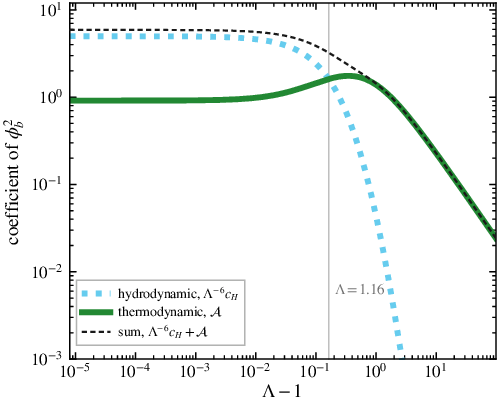}
  \caption{Coefficients of $\phi_b^{2}$ in the viscosity as a function of the excluded shell gap $\Lambda-1$. The hydrodynamic contribution $\Lambda^{-6}c_H$ (cyan, dotted) and the thermodynamic contribution $\mathcal{A}$ (green, solid) are shown together with their sum (black, dashed), which is the pair coefficient of $\eta_0/\mu$. The vertical line marks where the two contributions are
  equal, $\Lambda \approx 1.16$.}
  \label{fig:phi_b2_coefficients}
\end{figure}

The obtained amplitude can be compared with the pair coefficient of the
hydrodynamic viscosity, in terms of $\phi_b$ (Eq.~\eqref{eq:eta_inf_phi_b}).
Figure~\ref{fig:phi_b2_coefficients} shows both coefficients and their sum, which is the
pair coefficient of $\eta_0/\mu$. For hard spheres ($\Lambda=1$), the hydrodynamic coefficient is
more than five times greater than the thermodynamic one. The two become equal already at $\Lambda\simeq1.16$, while for thicker shells the
thermodynamic contribution dominates due to its scaling. For large $\Lambda$ the hydrodynamic coefficient falls as $\Lambda^{-6}$
(Eq.~\eqref{eq:eta_inf_phi_b}), and the thermodynamic one only as
$\Lambda^{-1}$ (Eq.~\eqref{eq:A_large_Lambda}).
This suggests that for particles with $\Lambda \gtrsim 1.16$
the hydrodynamic interactions are suppressed at a given $\phi_b$, and the pair
contribution to the zero-shear viscosity is dominated by the thermodynamic
stress.

\subsection{Thermodynamic viscosity at finite concentration}
\label{subsec:res_closures}

\begin{figure}[!tb]
  \panel{fig:eta_T_closures_a}{%
    \includegraphics[scale=1]{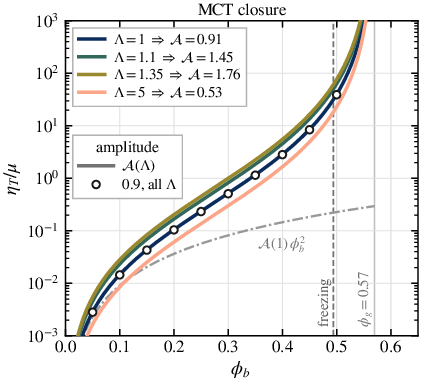}}\hfill
  \panel{fig:eta_T_closures_b}{%
    \includegraphics[scale=1]{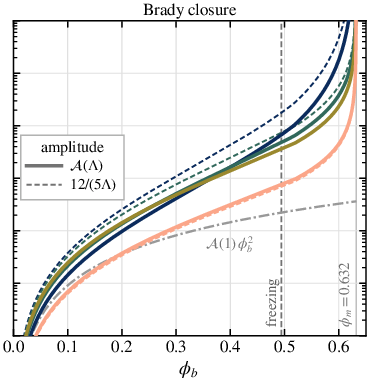}}
  \caption{Thermodynamic viscosity as a function of the thermodynamic volume fraction
  $\phi_b$, for $\Lambda=1$, $1.1$, $1.35$, and $5$.
  (a) The \gls{mct} closure. Thick curves use the exact amplitude, Eq.~\eqref{eq:mct}, while open circles show the result obtained using an amplitude of $0.9$, as in Russel \textit{et al.}~\cite{russelDivergenceLowShear2013a}, which gives the same curve for
  every $\Lambda$ and lies almost on the $\Lambda=1$ curve. (b) The Brady closure. Thick curves use the exact amplitude,
  Eq.~\eqref{eq:closure_ours}, while the thin dashed curves use the free-draining amplitude $12/(5\Lambda)$, Eq.~\eqref{eq:brady}.
  The slope discontinuity near $\phi_b=0.517$ comes from joining two branches of $g_c$, Eq.~\eqref{eq:pc}.
  In both panels the gray dash-dotted curve is the pair result for $\Lambda=1$,
  $\mathcal{A}(1)\phi_b^{2}$, the dashed vertical line marks hard-sphere
  freezing, and the solid vertical line indicates where the divergence occurs.}
  \label{fig:eta_T_closures}
\end{figure}

The pair result, Eq.~\eqref{eq:Adef}, fixes $\eta_T$ only to $\mathcal{O}(\phi_b^{2})$.
A closure is required to extend it to intermediate and high concentrations.
We use the amplitude-corrected \gls{mct} and Brady closures, Eqs.~\eqref{eq:mct} and~\eqref{eq:closure_ours}, with the amplitude $\mathcal{A}(\Lambda)$ obtained numerically from Eqs.~\eqref{eq:bvp}--\eqref{eq:AB}, and from here on we refer to them simply as the \gls{mct} and Brady closures.
Figure~\ref{fig:eta_T_closures} compares both with their original versions for several values of $\Lambda$.
The original closures have the same divergence factor, but keep the amplitude of their original formulation, $0.9$ for \gls{mct} and the free-draining form $12/(5\Lambda)$ for Brady.

As $\phi_b\to0$ every closure reduces to its amplitude times $\phi_b^{2}$, so only the
amplitude-corrected closures approach the exact pair result. The amplitude multiplies the divergence factor, so on the logarithmic axes of
Fig.~\ref{fig:eta_T_closures} changing it shifts a curve vertically by the
same amount at every $\phi_b$. In Fig.~\ref{fig:eta_T_closures_a} the original
\gls{mct} amplitude of $0.9$ is close to $\mathcal{A}(1)$, but it does not
depend on $\Lambda$. It therefore underestimates $\mathcal{A}$ for thin
shells and overestimates it for very thick ones.
In Fig.~\ref{fig:eta_T_closures_b} the original Brady closure lies above
the amplitude-corrected one for pure hard spheres and thin shells.
The two cross close to $\Lambda=1.35$, near the maximum of $\mathcal{A}$, and for thicker shells the original closure lies slightly below, until the two converge in the free-draining limit, $\Lambda \gg 1$.

The two divergence factors differ in how they depend on $\Lambda$ as well as
in how they diverge. The \gls{mct} factor depends on the structure alone. It is
therefore the same function of $\phi_b$ for every $\Lambda$, with the same
power-law divergence at $\phi_g$, and the curves of Fig.~\ref{fig:eta_T_closures_a} differ only
through the amplitude.
In contrast, the Brady factor also depends on $\Lambda$ through the short-time self-diffusivity, which is evaluated at the lower hydrodynamic volume fraction $\phi_a=\Lambda^{-3}\phi_b$.
A thicker shell therefore raises $\hat D^{\mathrm{s}}$ and lowers the Brady factor at fixed $\phi_b$.
The shell also changes the order of the pole at $\phi_m$. At $\Lambda=1$ both
$g_c$ and $1/\hat D^{\mathrm{s}}$ diverge there, and $\eta_T$ diverges as $(\phi_m-\phi_b)^{-2}$.
For $\Lambda>1$ the divergence of $g_c$ is unchanged, but $\hat D^{\mathrm{s}}$ stays finite
at $\phi_b=\phi_m$, so the divergence is only $(\phi_m-\phi_b)^{-1}$.

\begin{figure}[!tb]
  \panel{fig:eta_0_a}{%
    \includegraphics[scale=1]{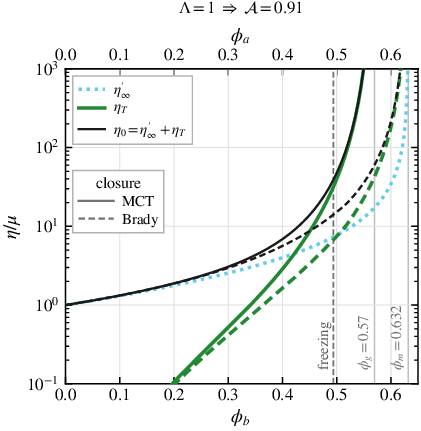}}\hfill
  \panel{fig:eta_0_b}{%
    \includegraphics[scale=1]{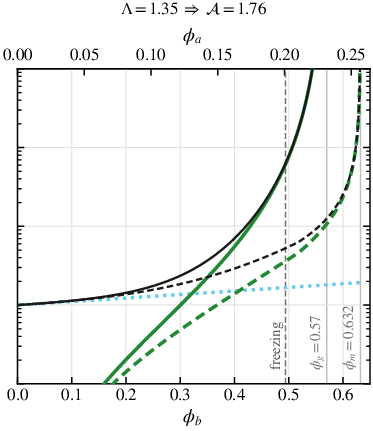}}
  \caption{Zero-shear viscosity $\eta_0=\eta'_\infty+\eta_T$ (black) and its
  hydrodynamic (cyan, dotted) and thermodynamic (green) contributions as a function of
  $\phi_b$. Panel (a) shows the results for pure hard spheres, while panel (b) gives the same functions at the maximum of
  $\mathcal{A}$ ($\Lambda=1.35$). The top axes give the hydrodynamic volume fraction
  $\phi_a=\Lambda^{-3}\phi_b$. Solid curves use the \gls{mct} closure,
  Eq.~\eqref{eq:mct}, and dashed curves the Brady closure,
  Eq.~\eqref{eq:closure_ours}, both with the exact amplitude. The hydrodynamic
  contribution, Eq.~\eqref{eq:eta_inf_phi_b}, is common to both. The dashed vertical
  line marks hard-sphere freezing, and the solid vertical lines the poles of the
  \gls{mct} and Brady divergence factors, at $\phi_g$ and $\phi_m$, respectively.}
  \label{fig:eta_0}
\end{figure}

Figure~\ref{fig:eta_0} shows how the two contributions combine in the
zero-shear viscosity. At low concentrations $\eta_0$ is dominated by
$\eta'_\infty$, and the two closures give nearly the same zero-shear
viscosity. For pure hard spheres $\eta'_\infty$ is still large near freezing,
and this common contribution reduces the difference between the closures.
At the maximum of $\mathcal{A}$ the shell keeps $\eta'_\infty$ small, so $\eta_0$ follows $\eta_T$, and the
closures differ by an order of magnitude at freezing.

\subsection{Comparison with experiment}
\label{subsec:res_experiment}

We finally compare the two-radius model with experimental measurements of the
zero-shear and high-frequency viscosity.
Sterically stabilized PMMA particles have become a standard experimental
hard-sphere system~\cite{puseyPhaseBehaviourConcentrated1986,royallSearchColloidalHard2013}.
A decisive test of the model, however, requires a suspension whose
hydrodynamic volume fraction is known independently of its rheology, and
ideally measurements of both $\eta_0$ and $\eta'_\infty$ on the same samples.
For sterically stabilized particles this is difficult. The grafted PHSA layer,
which is typically around $10$\,nm thick~\cite{bryantHowHardColloidal2002},
increases both radii by amounts that are not known separately.
The layer is also partly permeable to the
solvent, which may alter the hydrodynamic interactions between the particles%
~\cite{potaninHydrodynamicInteractionParticles1995}.
In addition, the volume fraction of such particles is difficult to determine,
and the available methods carry systematic errors of a few
percent~\cite{poonMeasuringColloidalVolume2012}, to which the steep rise of
$\eta_0$ at intermediate and high volume fractions is very sensitive.
None of this rules out PMMA data, and we return
to one such set below, but for a decisive test we prefer to focus on a suspension without a grafted layer.

\begin{figure*}[tb]
  \centering
  \includegraphics[scale=1]{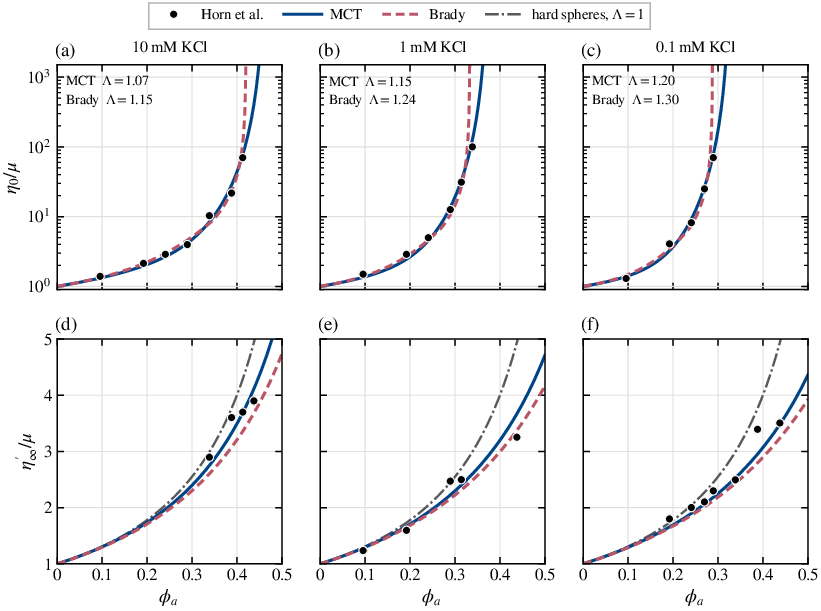}
  \caption{Zero-shear and high-frequency viscosity of the charge-stabilized
  polystyrene suspension PS200 ($a \approx 100$\,nm) of Horn
  \textit{et al.}~\cite{hornHydrodynamicColloidalInteractions2000} at three KCl
  concentrations. (a)--(c) Zero-shear viscosity. Solid blue curves use the
  \gls{mct} closure, Eq.~\eqref{eq:mct}, and dashed red curves the Brady closure,
  Eq.~\eqref{eq:closure_ours}, each with $\Lambda$ fitted separately at every
  salt concentration (values in the panels). (d)--(f) High-frequency
  viscosity, Eq.~\eqref{eq:eta_inf}, predicted with the same $\Lambda$ without
  further fitting. The gray dash-dotted curves in (d)--(f) are the hard-sphere prediction,
  $\Lambda=1$. The volume fraction
  $\phi_a$ is the one reported by Horn \textit{et al.}, obtained from the
  weight fraction and the density of polystyrene, and taken here as
  hydrodynamic.}
  \label{fig:horn-ps200}
\end{figure*}

Horn \textit{et al.}~\cite{hornHydrodynamicColloidalInteractions2000} used
such a suspension, namely a charge-stabilized polystyrene suspension. The particles have
no grafted polymer layer, and their volume fraction is determined directly from
the weight fraction, using the density of polystyrene. The reported volume
fraction should therefore be close to the hydrodynamic one.
Salt primarily affects the excluded-volume radius $b$, through the electrostatic repulsion~\cite{hornHydrodynamicColloidalInteractions2000}, and not the hydrodynamic radius $a$.
Horn \textit{et al.} measured
$\eta_0$ and $\eta'_\infty$ on the same samples. Figure~\ref{fig:horn-ps200} compares both with our theory, with $\Lambda$ fitted to $\eta_0$ alone, so that $\eta'_\infty$ is an independent test.

We fit a single $\Lambda$ at every salt concentration to their reported $\eta_0$ values
(Fig.~\ref{fig:horn-ps200}(a)--(c)). Both closures describe the data over the
whole range of volume fractions, and both give an excluded shell that thickens
as salt is removed and the electrostatic repulsion extends further. The \gls{mct}
closure gives thinner shells than the Brady closure, because its divergence at
$\phi_g$ is below $\phi_m$.

Horn \textit{et al.} analyzed the same data with an effective hard-sphere
volume fraction~\cite{hornHydrodynamicColloidalInteractions2000}. At every
salt concentration they extrapolated to the divergence of $\eta_0$ and rescaled
the volume fraction so that this divergence falls at $0.58$. The rescaled
zero-shear viscosities collapse onto a single master curve, and the
corresponding effective radius accounts for the excluded volume of the
electrostatic repulsion. Since both this rescaling and the \gls{mct} closure place
the divergence near the glass transition, their effective radii are close to
our \gls{mct} values of $\Lambda$. For the high-frequency viscosity, however,
Horn \textit{et al.} kept the measured rigid-core volume fraction. Their analysis
therefore already uses two volume fractions implicitly, an effective one for
the structure and the reported one for the hydrodynamics.

The two-radius model introduced here combines the two in one description.
Once $\Lambda$ is fitted to $\eta_0$, the theory predicts $\eta'_\infty$ at the reported volume fraction with no adjustable parameter.
The excluded shell keeps the particles apart and therefore
lowers $\eta'_\infty$ below the hard-sphere ($\Lambda=1$) prediction at the
same $\phi_a$ (Fig.~\ref{fig:horn-ps200}(d)--(f)). The data are noisy, but at
every salt concentration the densest samples lie below the hard-sphere curve,
as the shell requires. The \gls{mct} closure generally follows them more closely than
the Brady closure, whose thicker shells lower $\eta'_\infty$ too much.
Provided the reported volume fraction is the hydrodynamic one, the
high-frequency viscosity therefore favors a divergence of $\eta_0$ at the
glass transition rather than at random close packing, in agreement with Russel
\textit{et al.}~\cite{russelDivergenceLowShear2013a}.

We now turn to sterically stabilized PMMA particles, and focus on the data of Mewis \textit{et al.}~\cite{mewisRheologySuspensionsContaining1989}.
These data are a representative example of how an excluded shell around nearly hard spheres can be absorbed into an effective volume fraction, which accounts for the volume that the shell excludes, but not for the weaker hydrodynamic interactions between particles that the shell keeps apart.
Mewis \textit{et al.} obtained the volume fraction from the intrinsic viscosity of dilute suspensions, which corresponds to the hydrodynamic radius.
Against this hydrodynamic volume fraction, they found that the zero-shear viscosity of particles with a diameter of $475$\,nm lies above that of hard spheres~\cite{mewisRheologySuspensionsContaining1989}.
Russel \textit{et al.}~\cite{russelDivergenceLowShear2013a} later rescaled these
volume fractions by a factor $1.087$, and found that the rescaled
zero-shear data closely follow the `original' \gls{mct} closure with $\Lambda=1$.
However, the same correction can instead be read as an excluded shell of $\Lambda^{3}=1.087$
around particles whose hydrodynamic volume fraction is the reported one.
Both readings then give the same $\phi_b$, and they differ only in $\phi_a$.
With this shell and nothing else fitted, the \gls{mct} closure also follows the data
well, and the two descriptions seem practically indistinguishable
(Fig.~\ref{fig:mewis_a}). Brady's closure with the same shell instead falls
well below the data as the divergence approaches, and would need a thicker
shell to fit them, consistent with the conclusion that we reached for the
suspensions of Horn \textit{et al.}~\cite{hornHydrodynamicColloidalInteractions2000}.

\begin{figure}[!tb]
  \panel{fig:mewis_a}{%
    \includegraphics[scale=1]{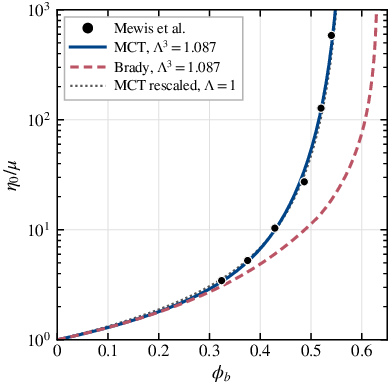}}\hfill
  \panel{fig:mewis_b}{%
    \includegraphics[scale=1]{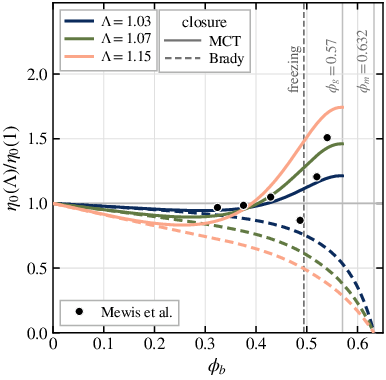}}
  \caption{(a) Zero-shear viscosity of the PMMA particles of Mewis
  \textit{et al.}~\cite{mewisRheologySuspensionsContaining1989} against
  $\phi_b=1.087\,\phi$, where $\phi$ is their volume fraction from the intrinsic
  viscosity. \Gls{mct} (solid blue) and Brady (dashed red) closures with
  $\Lambda^{3}=1.087$, which take $\phi$ as $\phi_a$.
  We also show the result for hard spheres with $\Lambda=1$ in the \gls{mct} closure (dotted), where $\phi$ is rescaled.
  Here, rescaled means that, as in Russel
  \textit{et al.}~\cite{russelDivergenceLowShear2013a}, the factor $1.087$
  corrects the volume fraction itself, so that $\phi_a=\phi_b=1.087\,\phi$. (b) Zero-shear viscosity of the two-radius model relative
  to that of hard spheres at the same $\phi_b$, for the \gls{mct} (solid) and Brady
  (dashed) closures. Black circles show the viscosities from panel (a) divided
  by the \gls{mct} hard-sphere viscosity at the same $\phi_b$ for $\Lambda=1$, with the exact amplitude $\mathcal{A}(1)$.
  Curves are shown for $\Lambda=\sqrt[3]{1.087}\approx1.03$, the shell of panel (a), and for the thicker shells $\Lambda=1.07$ and $1.15$. Vertical
  lines mark freezing
  (dashed) and the poles $\phi_g$ and $\phi_m$ (solid).}
  \label{fig:mewis}
\end{figure}

This near-equivalence of the two interpretations holds generally for thin shells
in the \gls{mct} closure (Fig.~\ref{fig:mewis_b}). At fixed $\phi_b$ its divergence factor is the same
for every $\Lambda$, so the two-radius model and hard spheres at the same
$\phi_b$ differ only in the amplitude and in $\eta'_\infty$.
At intermediate concentrations $\eta'_\infty$ is still a large part of $\eta_0$, so the lower $\eta'_\infty$ dominates and the ratio of the two viscosities falls below unity.
At higher concentrations $\eta_T$ takes over, so the amplitude dominates and the ratio rises above unity until it tends to $\mathcal{A}(\Lambda)/\mathcal{A}(1)$ at the divergence.
For thin shells the ratio therefore stays close to unity, and the two interpretations
are then practically indistinguishable throughout the concentration range,
whereas for thicker shells the two descriptions separate.
Brady's closure has no such equivalence even for thin shells, because its
divergence factor depends on $\phi_a$ through the short-time self-diffusivity. At
the same $\phi_b$ the shell lowers $\phi_a$, so the diffusivity remains finite
at $\phi_m$ and the divergence is weaker than that expected for hard spheres. This causes the ratio to fall to zero at $\phi_m$.

In order to examine the two readings of the experimental data more closely, we
divide the measured viscosities of Mewis \textit{et al.} by the \gls{mct} hard-sphere
viscosity at the same $\phi_b=1.087\,\phi$ (black circles in
Fig.~\ref{fig:mewis_b}). If the factor $1.087$ only corrects the volume fraction,
the points should lie at unity, whereas if it instead reflects an excluded shell, they
should follow the solid curve for $\Lambda=\sqrt[3]{1.087}\approx1.03$.
The measured ratios scatter around this expected curve.
The two most dilute samples lie slightly below unity, as the curve does, and the
two densest samples lie above it.
The trend follows the shell curve, but the deviations from unity are small
enough that errors in the volume fraction could also explain them. Close to the divergence, $\eta_0$ is so steep that
even a small error in the volume fraction, whether in the original measurement
or in digitization, changes the ratio considerably.
The data are therefore indicative of a thin shell, but should not be taken to be conclusive. For a
decisive test, more accurate measurements of the absolute hydrodynamic
volume fraction and of $\eta_0$ are required.

Both experimental systems suggest that an excluded shell may be present in
electrostatically and sterically stabilized particles. 
Earlier work on such suspensions that mapped their zero-shear viscosity or freezing
onto hard spheres through a single effective volume fraction%
~\cite{puseyPhaseBehaviourConcentrated1986, segreViscosityStructuralRelaxation1995, phanPhaseTransitionEquation1996, meekerConcentrationDependenceLowshear1997, russelDivergenceLowShear2013a}
may therefore be partly affected by the presence of a shell.
We want to be clear that this does not exclude the possibility of errors in the absolute volume fraction%
~\cite{poonMeasuringColloidalVolume2012}; a data set may contain both effects.
We suggest that future experiments on nearly hard-sphere suspensions treat the two radii as separate (fit) parameters using the effective descriptions put forward in this work and report the hydrodynamic volume fraction independently of the zero-shear viscosity.
The same recommendation applies to measurements of the linear viscoelastic spectrum%
~\cite{swanMediumAmplitudeOscillatory2014} between $\eta'_\infty$ and
$\eta_0$, and of nonlinear rheology~\cite{bergenholtzNonNewtonianRheologyDilute2002}.

Turning to a more practical viewpoint, the shell could be used to tune the rheology of a suspension, because it partially filters the hydrodynamic interactions between particles.
It keeps pairs out of surface gaps smaller than $2(b-a)$, where lubrication dominates, and leaves far-field interactions unchanged.
At fixed $\phi_b$, a thicker shell therefore shifts the zero-shear viscosity from hydrodynamic to thermodynamic stress (Figs.~\ref{fig:phi_b2_coefficients} and~\ref{fig:eta_0}).
At high shear rates, Bergenholtz \textit{et al.}~\cite{bergenholtzNonNewtonianRheologyDilute2002} found that shells thicker than $\Lambda\simeq1.1$ suppress the shear thickening caused by lubrication, so that the viscosity reaches a Newtonian high-shear plateau.
Experimentally, the shell thickness can be tuned, for example by grafting polymer layers of different thicknesses onto the particle surface~\cite{mewisRheologySuspensionsContaining1989}, or by using electrostatic repulsion to create an effective shell~\cite{hornHydrodynamicColloidalInteractions2000}.

The two-radius model may also have value in describing phenomena outside the realm of
rheology in which thermodynamic and hydrodynamic interactions are both relevant.
Examples include short-time diffusion and
sedimentation~\cite{riestDynamicsSuspensionsHydrodynamically2015}; long-time
self-diffusion~\cite{cichockiDiffusionBrownianParticles1991}; and crystal
nucleation~\cite{kurtenNarrowingCauseHardsphere2026}, whose rate is highly sensitive to the volume
fraction~\cite{poonMeasuringColloidalVolume2012}.

%% file: sections/conclusions.tex
\section{Conclusion and Outlook}
\label{sec:conclusions}

Summarizing, we have developed a theory for the zero-shear and high-frequency viscosity of hard spheres whose excluded-volume radius $b$ is larger than their hydrodynamic radius $a$.
The theory splits the zero-shear viscosity into a hydrodynamic part, equal to the high-frequency viscosity $\eta'_\infty$, and a thermodynamic part $\eta_T$, Eq.~\eqref{eq:zero-shear_viscosity_total}.
We calculated both parts exactly at the pair level, and we extended them to finite volume fractions with closures,
without any adjustable parameters apart from the shell thickness itself.
A self-consistent closure gives $\eta'_\infty$, Eq.~\eqref{eq:eta_inf}, and two alternative closures give $\eta_T$, Eqs.~\eqref{eq:mct} and~\eqref{eq:closure_ours}.
These final expressions are analytical, except for two pair-level functions of $\Lambda=b/a$: the integral $I(\Lambda)$ of the pair stresslet in the hydrodynamic part and the amplitude $\mathcal{A}(\Lambda)$ in the thermodynamic part.
Both functions follow from the full two-sphere hydrodynamics, which we evaluated numerically for every $\Lambda$, and we fitted them with semi-empirical expressions, Eqs.~\eqref{eq:I_approx} and~\eqref{eq:A_approx}.
With these expressions the full model becomes analytical, so that both viscosities can be evaluated at any shell thickness and volume fraction, and compared with experiments and/or simulations, without the need for the two-sphere calculations.

The presence of a shell affects the two parts of the viscosity in different ways. (i)~In the hydrodynamic part, it removes the pair configurations closest to contact, where the hydrodynamic interactions are strongest.
It therefore lowers $\eta'_\infty$ and keeps it finite even when the excluded-volume spheres reach close packing, where $\eta'_\infty$ of hard spheres diverges.
Our \gls{sd} simulations follow the self-consistent closure up to freezing (Fig.~\ref{fig:eta_inf_theory}). (ii)~In the thermodynamic part, the effect of the shell at pair level is nonmonotonic (Fig.~\ref{fig:A_of_x_b}).
Lubrication prevents the flow from pushing hard spheres into contact, but with a shell the flow can push pairs until their shells touch, so that contact forces add to the thermodynamic stress.
The amplitude of $\eta_T$ therefore rises to nearly twice its hard-sphere value near $\Lambda=1.35$, before it decays toward the free-draining limit for thicker shells.

Of the two closures for $\eta_T$ considered in this work, the first, based on \gls{mct}~\cite{russelDivergenceLowShear2013a}, diverges at the glass transition.
The second, based on Brady's model~\cite{bradyRheologicalBehaviorConcentrated1993}, diverges at random close packing.
In their original form, the \gls{mct} closure has the right amplitude only for hard spheres, and Brady's closure only for thick shells.
We replaced both amplitudes by the exact pair amplitude, so that the closures apply at any shell thickness.
In both closures, a thicker shell at fixed $\phi_b$ shifts the zero-shear viscosity from hydrodynamic to thermodynamic stress (Figs.~\ref{fig:phi_b2_coefficients} and~\ref{fig:eta_0}).

The comparison with experiments indicates that our complete two-radius theory can describe both $\eta'_\infty$ and $\eta_0$.
For the charge-stabilized polystyrene particles of Horn \textit{et al.}~\cite{hornHydrodynamicColloidalInteractions2000}, the shell fitted to $\eta_0$ thickens as salt is removed, and it lowers $\eta'_\infty$ below the hard-sphere curve, consistent with the experimentally measured values (Fig.~\ref{fig:horn-ps200}).
The \gls{mct} closure follows these data more closely, which favors a divergence of $\eta_0$ at the glass transition, provided the reported volume fraction is the hydrodynamic one.
For the sterically stabilized PMMA particles of Mewis \textit{et al.}~\cite{mewisRheologySuspensionsContaining1989}, a thin shell and a rescaled volume fraction describe $\eta_0$ equally well in the \gls{mct} closure (Fig.~\ref{fig:mewis}).
These data are indicative of a thin shell, and more precise measurements of the hydrodynamic volume fraction, together with $\eta_0$ or $\eta'_\infty$, would show decisively whether a shell is present and how thick it is.

Looking forward, we have provided the analytic expressions needed for simulation to validate the predictions of, and for experiment to test the applicability of, the two-radius model and its closures.
Brownian \gls{sd} simulations can be used to contrast the two closures for $\eta_T$ at higher volume fractions.
Accurately reproducing hard-sphere dynamics near shell contact is essential but numerically challenging, as particle trajectories can depend sensitively on the time integration and the treatment of close encounters~\cite{ekiel-jezewskaStokesianDynamicsClose2008, bromsBarrierMethodContact2024}.
Repeating our simulations of $\eta'_\infty$ with polydisperse particles could suppress crystallization and allow comparisons with the theory and experiments beyond freezing.
In experiments, precise measurements of $\eta'_\infty$ could be used to
determine the shell thickness of real suspensions, and the measured $\eta_0$ could then be compared with the prediction at the same shell thickness. As indicated, this can be relevant to better understand the behavior of dense nearly hard-sphere suspensions. In particular, we encourage the adoption of the two-radius model in systems where thermodynamic and hydrodynamic interactions are both relevant, such as those found in sedimentation, diffusion, and potentially crystal nucleation.

It is known that particles with internal hydrodynamic structure can be represented approximately within the two-radius model by an effective hydrodynamic radius~\cite{cichockiHydrodynamicRadiusApproximation2014}. This suggests it is possible to describe a wider range of colloidal suspensions. Our treatment could also be extended to soft repulsive and attractive interactions by incorporating the interaction potential into the pair Smoluchowski equation, equilibrium pair distribution and stress integrals~\cite{vargaLinearViscoelasticityAttractive2015}.

%% file: sections/acknowledgments.tex
\begin{acknowledgments}
We thank Stefan Schmid for useful discussions on the high-frequency viscosity.

This work is part of the CoCoGel project, which has received funding from
the European Union's Horizon Europe research and innovation programme under
the Marie Sk\l{}odowska-Curie grant agreement No.~101120301.

Views and opinions expressed are however those of the authors only and do
not necessarily reflect those of the European Union or the European
Research Executive Agency (REA). Neither the European Union nor the
granting authority can be held responsible for them.

We also acknowledge funding from the Dutch Research Council (NWO) under grant OCENW.KLEIN.354.

Generative AI was used to edit parts of the text and to help make the numerical solution of the pair Smoluchowski problem and the calculation of the hydrodynamic functions efficient.
The authors verified and edited all text and code produced with this assistance and take full responsibility for the content of this article.
\end{acknowledgments}

%% file: sections/declarations.tex
\section*{Author Declarations}

\subsection*{Conflict of Interest}
The authors have no conflicts to disclose.

\subsection*{Author Contributions}
\textbf{Dimos Aslanis}: Conceptualization; Methodology; Software; Formal analysis; Investigation; Visualization; Writing -- original draft.

\textbf{Joost de Graaf}: Conceptualization; Methodology; Supervision; Funding acquisition; Writing -- review \& editing.

\section*{Data Availability}
The code that computes the pair functions, the closures and the figures, and the Stokesian dynamics data for the high-frequency viscosity, will be made openly available in a public repository upon publication.

%% file: sections/appendix.tex
\clearpage
\setcounter{equation}{0}
\renewcommand{\theequation}{A\arabic{equation}}

\section*{Appendix}
\setcounter{subsection}{0}
\renewcommand{\thesubsection}{\Alph{subsection}}
\makeatletter\def\p@subsection{}\makeatother

\subsection{Pair-level thermodynamic stresses}
\label{app:pair-stress}

We derive Eqs.~\eqref{eq:AP} and \eqref{eq:AB} following
Ref.~\onlinecite{russelHugginsCoefficientMeans1984}, keeping $\Lambda$ explicit.
At pair level, the stresses can be written in terms of the functions $A$ and
$B$ of
Eq.~\eqref{eq:Urel}~\cite{batchelorEffectBrownianMotion1977,russelHugginsCoefficientMeans1984}:
\begin{align}
  n\langle\mathbf{S}^{P}\rangle
  &= \mathrm{I.T.}+\frac{n^{2}}{2}\int r\bigl[1-A(r/a)\bigr]\hat{\mathbf{r}}\hat{\mathbf{r}}\,
     V'(r)\,g(\mathbf{r})\,\mathrm{d}\mathbf{r},
  \label{eq:app-SP-pair}\\
  n\langle\mathbf{S}^{B}\rangle
  &= \mathrm{I.T.}-\frac{n^{2}k_BT}{2}\int W(r/a)\,\hat{\mathbf{r}}\hat{\mathbf{r}}\,
     g(\mathbf{r})\,\mathrm{d}\mathbf{r},
  \label{eq:app-SB-pair}
\end{align}
where $V(r)$ is the pair potential of Eq.~\eqref{eq:V_shell} and
$\mathrm{I.T.}$ denotes isotropic
terms~\cite{bradyBrownianMotionHydrodynamics1993,bergenholtzNonNewtonianRheologyDilute2002},
which do not affect the viscosity.
Both pair stresses are proportional to $n^2$, so their $\mathcal{O}(\phi_b^2)$
contributions require only the dilute-limit pair distribution.

For the excluded-shell potential, Eq.~\eqref{eq:V_shell}, the dilute-limit
equilibrium pair distribution is given by Eq.~\eqref{eq:pair_eq}, so that
$V'(r)\,g_{\mathrm{eq}}(r)=-k_BT\,\delta(r-2b)$, which reduces the volume
integral in Eq.~\eqref{eq:app-SP-pair} to the contact sphere $r = 2b$.
In Eq.~\eqref{eq:app-SB-pair}, $g_{\mathrm{eq}}(r)=\Theta(r-2b)$ restricts
the radial integral to $r\geq2b$. Substituting Eq.~\eqref{eq:distortion} into
both expressions, we obtain
\begin{align}
  n\langle\mathbf{S}^{P}\rangle
  &= \mathrm{I.T.}-4n^{2}b^{3}k_BT\bigl[1-A(2\Lambda)\bigr]\Bigl[
     \oint\hat{\mathbf{r}}\hat{\mathbf{r}}\,\mathrm{d}\Omega
     \nonumber\\
  &\qquad
     -\mathrm{Pe}_b\,f(2)\oint\hat{\mathbf{r}}\hat{\mathbf{r}}\,\bigl(\hat{\mathbf{r}}\cdot\hat{\mathbf{E}}\cdot\hat{\mathbf{r}}\bigr)\,\mathrm{d}\Omega
     \Bigr],
  \label{eq:app-SP-expanded}\\
  n\langle\mathbf{S}^{B}\rangle
  &= \mathrm{I.T.}-\frac{n^{2}k_BT}{2}\int_{2b}^{\infty}W(r/a)\Bigl[
     \oint\hat{\mathbf{r}}\hat{\mathbf{r}}\,\mathrm{d}\Omega
     \nonumber\\
  &\qquad
     -\mathrm{Pe}_b\,f(r/b)\oint\hat{\mathbf{r}}\hat{\mathbf{r}}\,\bigl(\hat{\mathbf{r}}\cdot\hat{\mathbf{E}}\cdot\hat{\mathbf{r}}\bigr)\,\mathrm{d}\Omega
     \Bigr]r^{2}\,\mathrm{d}r.
  \label{eq:app-SB-expanded}
\end{align}
The angular integrals are
\begin{equation}
  \oint\hat{\mathbf{r}}\hat{\mathbf{r}}\,\mathrm{d}\Omega=\frac{4\pi}{3}\,\mathbf{I},
  \qquad
  \oint\hat{\mathbf{r}}\hat{\mathbf{r}}\,\bigl(\hat{\mathbf{r}}\cdot\hat{\mathbf{E}}\cdot\hat{\mathbf{r}}\bigr)\,\mathrm{d}\Omega
  =\frac{8\pi}{15}\,\hat{\mathbf{E}} ,
  \label{eq:app-angular}
\end{equation}
where $\hat{\mathbf{E}}$ is traceless because the flow is incompressible.
With $r=b\rho$, Eqs.~\eqref{eq:app-SP-expanded} and \eqref{eq:app-SB-expanded}
then become
\begin{align}
  n\langle\mathbf{S}^{P}\rangle
  &= \mathrm{I.T.}+\frac{32\pi}{15}\,n^{2}b^{3}k_BT\,\mathrm{Pe}_b
     \left[1-A(2\Lambda)\right]f(2)\,\hat{\mathbf{E}},
  \label{eq:app-SP-dev}\\
  n\langle\mathbf{S}^{B}\rangle
  &= \mathrm{I.T.}+\frac{4\pi}{15}\,n^{2}b^{3}k_BT\,\mathrm{Pe}_b
     \int_{2}^{\infty}W(\Lambda\rho)\,f(\rho)\,\rho^{2}\,\mathrm{d}\rho\;
     \hat{\mathbf{E}}.
  \label{eq:app-SB-dev}
\end{align}
Writing the deviatoric stresses in the Newtonian forms
$2\eta_P\langle\mathbf{E}\rangle$ and $2\eta_B\langle\mathbf{E}\rangle$, with
$\langle\mathbf{E}\rangle=\dot\gamma\hat{\mathbf{E}}$, and using
$n^{2}b^{3}k_BT\,\mathrm{Pe}_b=27\mu\dot\gamma\,\phi_b^{2}/(16\pi\Lambda)$, and
$n=3\phi_b/(4\pi b^{3})$, gives Eqs.~\eqref{eq:AP} and \eqref{eq:AB}.

\subsection{Numerical evaluation of $\mathcal{A}(\Lambda)$ and $J(s)$}
\label{app:numerical-method}

This subsection describes the evaluation of the pair amplitude
\(\mathcal{A}=\mathcal{A}_P+\mathcal{A}_B\) of Eqs.~\eqref{eq:Adef},
\eqref{eq:AP}, and \eqref{eq:AB} for \(\Lambda=b/a>1\), and of the
kernel $J(s)$ of Eq.~\eqref{eq:J_mobility}.

Following Ref.~\onlinecite{bergenholtzNonNewtonianRheologyDilute2002},
we map the domain $2\leq\rho<\infty$ of Eqs.~\eqref{eq:bvp} and \eqref{eq:bvpbc} onto
$0<t\leq1$ with
\begin{equation}
    t=\frac{2}{\rho},
    \qquad
    s=\frac{2\Lambda}{t},
    \label{eq:app-coordinate-transform}
\end{equation}
so that $t=1$ is excluded-volume contact, where $s=2\Lambda$, and
$t=0$ is infinite separation.  The function $F(t)=f(2/t)$ then obeys
\begin{equation}
    G\,t^2F''-2\Lambda G_sF'-6HF=-\frac{4W}{t^2},
    \label{eq:app-bvp}
\end{equation}
where $G_s=\mathrm{d}G/\mathrm{d}s$ and the hydrodynamic functions are evaluated at
$s=2\Lambda/t$, subject to
\begin{equation}
    F(0)=0,
    \qquad
    F'(1)=q_c\equiv4\,\frac{1-A(2\Lambda)}{G(2\Lambda)}.
    \label{eq:app-boundary-conditions}
\end{equation}
We evaluate $W$ using the equivalent numerically stable form%
~\cite{cichockiLongtimeSelfdiffusionCoefficient1988, batchelorEffectBrownianMotion1977}
\begin{equation}
    W(s)=3B(s)-\frac{1}{s^2}\frac{\mathrm{d}P_A}{\mathrm{d}s},
    \qquad
    P_A\equiv s^3A.
    \label{eq:app-W-stable}
\end{equation}

For equal spheres, the functions of Eqs.~\eqref{eq:Drel} and
\eqref{eq:Urel} are relative-motion combinations of the two-sphere
mobility functions of Kim and
Mifflin~\cite{kimResistanceMobilityFunctions1985},
\begin{align}
    G&=x^a_{11}-x^a_{12},
    &A&=\frac{4}{s}\bigl(x^g_{11}-x^g_{12}\bigr),
    \nonumber\\
    H&=y^a_{11}-y^a_{12},
    &B&=\frac{8}{s}\bigl(y^g_{11}-y^g_{12}\bigr).
    \label{eq:app-mobility-definitions} 
\end{align}
The near-contact asymptotics are known for the resistance
functions~\cite{jeffreyCalculationResistanceMobility1984,
jeffreyCalculationLowReynolds1992,townsendGeneratingScratchNearfield2023} instead. 
For a pair, $\boldsymbol{\mathcal{R}}$ in Eq.~\eqref{eq:GRM} is expressed
through scalar resistance functions of $s$, such as $X^K_{\alpha\beta}$ and
$Y^K_{\alpha\beta}$, where $\alpha,\beta=1,2$ label the spheres, $K$ labels
the coupling between force, torque, or stresslet and translation, rotation,
or strain, and $X$ and $Y$ denote axisymmetric and transverse
components~\cite{jeffreyCalculationResistanceMobility1984,kimResistanceMobilityFunctions1985}.  We use the
notation and normalization of
Ref.~\onlinecite{kimMicrohydrodynamicsPrinciplesSelected2005} for both resistance and mobility
functions. The two sets are related by matrix inversion~\cite{kimResistanceMobilityFunctions1985}. 
For equal spheres, it can be reduced to
combinations of $\Delta Z=Z_{11}-Z_{12}$ and $\Sigma Z=Z_{11}+Z_{12}$.  With
$D\equiv3\Delta Y^A\Sigma Y^C-(\Delta Y^B)^2$,
Eq.~\eqref{eq:app-mobility-definitions} becomes
\begin{align}
    G&=\frac{1}{\Delta X^A},
    &A&=\frac{4}{3s}\frac{\Delta X^G}{\Delta X^A},
    \nonumber\\
    H&=\frac{3\Sigma Y^C}{D},
    &B&=\frac{8}{s}\,
    \frac{\Delta Y^G\Sigma Y^C+\Sigma Y^H\Delta Y^B}{D}.
    \label{eq:app-resistance-form}
\end{align}

Following Ref.~\onlinecite{townsendMechanicsSuspensions2017}, we treat
near and far separations separately.
For gaps $\xi=s-2\leq10^{-2}$, we take the resistance functions from the
corrected~\cite{townsendGeneratingScratchNearfield2023} lubrication expressions~\cite{jeffreyCalculationResistanceMobility1984,
jeffreyCalculationLowReynolds1992}.
For $\xi\geq2\times10^{-2}$, we compute the mobility functions with the
multipole-reflection method~\cite{wilsonStokesFlowThree2013} and
invert them to obtain the resistance
functions~\cite{kimResistanceMobilityFunctions1985}.  In between, the two
are joined by monotone cubic interpolation in $\log\xi$.
In every regime, $G$, $H$, $A$, and $B$ follow from the resistance functions
through Eqs.~\eqref{eq:app-resistance-form}.  For
$\xi<2\times10^{-2}$, these are evaluated in closed form at each grid point.
The multipole-reflection method is too costly for that, so for
$\xi\geq2\times10^{-2}$ we evaluate them only at selected separations and
represent $G$, $H$, $P_A$, and $s^5B$ in between by Chebyshev interpolants
in $\log\xi$ for $s<30$, and by power series%
~\cite{batchelorBrownianDiffusionParticles1976,
batchelorHydrodynamicInteractionTwo1972} in $1/s$ for $s\geq30$.
In each regime, $G_s$ and $\mathrm{d}P_A/\mathrm{d}s$ are found analytically.

Equation~\eqref{eq:app-bvp} is solved by finite differences.
The contact term of Eq.~\eqref{eq:AP} then follows from $f(2)=F(1)$,
and the bulk term from Eq.~\eqref{eq:AB} written in $t$,
\begin{equation}
    \mathcal{A}_{B}(\Lambda)
    =\frac{9}{5\Lambda}
      \int_0^1\frac{W(2\Lambda/t)F(t)}{t^4}\,\mathrm{d}t.
    \label{eq:app-AB}
\end{equation}

The kernel $J(s)$ of Eq.~\eqref{eq:J_mobility} is computed from the same
resistance functions. Expressing the mobility functions through the
resistance functions, Eq.~\eqref{eq:J_mobility} becomes
\begin{equation}
  \begin{aligned}
    J(s) = -1 + \frac{1}{5}\biggl[\,
      &\Sigma X^M - \frac{4}{15}\,\frac{(\Delta X^G)^2}{\Delta X^A}
      + 2\,\Sigma Y^M + 2\,\Sigma Z^M \\
      &- \frac{24}{5}\,
      \frac{\Sigma Y^C(\Delta Y^G)^2
        + 2\,\Delta Y^B\,\Delta Y^G\,\Sigma Y^H
        + 3\,\Delta Y^A(\Sigma Y^H)^2}{D}
    \biggr].
  \end{aligned}
  \label{eq:app-J-resistance}
\end{equation}
For the calculation of $J(s)$, we use the same near-field and far-field methods
as for the hydrodynamic functions. For $\xi<10^{-7}$ we use its asymptotic form,
Eq.~\eqref{eq:J_contact}.

\subsection{Large-$\Lambda$ expansion of the pair amplitude}
\label{app:large-Lambda}

For $\Lambda\gg1$ the excluded shell keeps every pair at $s=\Lambda\rho\ge2\Lambda$,
far outside the lubrication region. We therefore use the far-field forms of the
hydrodynamic
functions~\cite{batchelorBrownianDiffusionParticles1976,batchelorEffectBrownianMotion1977}.
Evaluated at $s=\Lambda\rho$, each power $1/s$ becomes $\epsilon/\rho$, with
$\epsilon=\Lambda^{-1}$, and
\begin{equation}
\begin{aligned}
  G(\Lambda\rho)&\simeq1-\frac{3\epsilon}{2\rho}+\frac{\epsilon^{3}}{\rho^{3}}, &
  H(\Lambda\rho)&\simeq1-\frac{3\epsilon}{4\rho}-\frac{\epsilon^{3}}{2\rho^{3}},\\
  A(\Lambda\rho)&\simeq\frac{5\epsilon^{3}}{\rho^{3}}, &
  W(\Lambda\rho)&\simeq\frac{75\epsilon^{6}}{\rho^{6}}.
\end{aligned}
  \label{eq:app-far-field}
\end{equation}
We substitute $f=\sum_{n}\epsilon^{n}f_n$ into
Eqs.~\eqref{eq:bvp} and \eqref{eq:bvpbc} and collect powers of $\epsilon$. Since
$W$ is of order $\epsilon^{6}$, it drops out of Eq.~\eqref{eq:bvp} up to order
$\epsilon^{3}$. At leading order $G=H=1$, and the problem
reduces to the free-draining one,
\begin{equation}
  f_0''+\frac{2}{\rho}f_0'-\frac{6}{\rho^{2}}f_0=0, \qquad f_0'(2)=-2,
\end{equation}
with the decaying solution~\cite{russelHugginsCoefficientMeans1984,bradyRheologicalBehaviorConcentrated1993}
$f_0=32/(3\rho^{3})$.
At order $\epsilon$, the terms linear in $\epsilon$ in $G$ and $H$ act on
$f_0$, and
\begin{equation}
  f_1''+\frac{2}{\rho}f_1'-\frac{6}{\rho^{2}}f_1=\frac{96}{\rho^{6}}, \qquad
  f_1'(2)=-\frac{3}{2},
\end{equation}
with the decaying solution $f_1=8(6-\rho)/(3\rho^{4})$. The orders
$\epsilon^{2}$ and $\epsilon^{3}$ follow in the same way, giving
\begin{align*}
  f_2(\rho)&=\frac{-13\rho^2-84\rho+468}{21\rho^5}, &
  f_3(\rho)&=\frac{-1045\rho^3-156\rho^2-936\rho+3804}{168\rho^6},
\end{align*}
and hence
\begin{equation}
  f(2)=\frac{4}{3}+\frac{2}{3\Lambda}+\frac{31}{84\Lambda^{2}}
    -\frac{1763}{2688\Lambda^{3}}+\mathcal{O}(\Lambda^{-4}).
  \label{eq:app-f2}
\end{equation}
Inserting Eq.~\eqref{eq:app-f2} and
$1-A(2\Lambda)=1-5/(8\Lambda^{3})+\mathcal{O}(\Lambda^{-5})$ into Eq.~\eqref{eq:AP}, and
$W$ and $f_0$ into Eq.~\eqref{eq:AB}, gives Eq.~\eqref{eq:A_large_Lambda}.